\documentclass[english]{IEEEtran}
\usepackage{graphicx}
\usepackage{amssymb,amsbsy}
\usepackage{epstopdf}
\usepackage{amsmath,amsthm}
\usepackage{enumerate}
\usepackage{eufrak}
\usepackage{cite}
\usepackage{mathcomp}
\usepackage{supertabular}
\usepackage{longtable}
\usepackage{stmaryrd}
\usepackage{url}
\usepackage{color}
\usepackage{rotating}
\usepackage{float}

\usepackage{array}
\usepackage{mathtools}
\usepackage{multicol}
\usepackage{amsmath}
\usepackage{algorithm}
\usepackage{algorithmic}
\usepackage{amssymb}
\usepackage{xcolor}

\usepackage[all]{xy}
\entrymodifiers={++[o][F-]}

\newcommand{\I}{\mathcal I}

\newcommand{\enc}{\textsc{Encode}}
\newcommand{\dec}{\textsc{Decode}}

\theoremstyle{definition}

\newtheorem{prop}{Proposition}[section]
\newtheorem{thm}[prop]{Theorem}

\newtheorem{defn}{Definition}[section]

\usepackage[T1]{fontenc}
\usepackage[latin9]{inputenc}
\usepackage{amsmath}
\usepackage{graphicx}
\usepackage{color}
\usepackage{multicol}

\makeatletter
\numberwithin{equation}{section}
\numberwithin{figure}{section}

\makeatother

\usepackage{babel}

\makeatletter
\numberwithin{equation}{section}
\numberwithin{figure}{section}

\makeatother

\usepackage{babel}

\begin{document}

\newcommand{\vA}{{\bf A}}
\newcommand{\vAtilde}{\widetilde{\bf A}}

\newcommand{\vB}{{\bf B}}
\newcommand{\vBtilde}{\widetilde{\bf B}}

\newcommand{\vC}{{\bf C}}
\newcommand{\vD}{{\bf D}}
\newcommand{\vH}{{\bf H}}
\newcommand{\vI}{{\bf I}}

\newcommand{\vY}{{\bf Y}}
\newcommand{\vZ}{{\bf Z}}

\newcommand{\vJ}{{\bf J}}

\newcommand{\vM}{{\bf M}}
\newcommand{\vN}{{\bf N}}
\newcommand{\vU}{{\bf U}}
\newcommand{\vV}{{\bf V}}
\newcommand{\vT}{{\bf T}}
\newcommand{\vR}{{\bf R}}
\newcommand{\vS}{{\bf S}}

\newcommand{\va}{{\bf a}}
\newcommand{\vb}{{\bf b}}
\newcommand{\vc}{{\bf c}}

\newcommand{\ve}{{\bf e}}
\newcommand{\vh}{{\bf h}}
\newcommand{\vp}{{\bf p}}

\newcommand{\vu}{{\bf u}}
\newcommand{\vv}{{\bf v}}
\newcommand{\vw}{{\bf w}}
\newcommand{\vx}{{\bf x}}
\newcommand{\vhx}{{\widehat{\bf x}}}
\newcommand{\vtx}{{\widetilde{\bf x}}}
\newcommand{\vy}{{\bf y}}
\newcommand{\vz}{{\bf z}}

\newcommand{\vj}{{\bf j}}
\newcommand{\vzero}{{\bf 0}}
\newcommand{\vone}{{\bf 1}}
\newcommand{\vbeta}{{\boldsymbol \beta}}
\newcommand{\vchi}{{\boldsymbol \chi}}

\newcommand{\dA}{\mathtt A}
\newcommand{\dT}{\mathtt T}
\newcommand{\dC}{\mathtt C}
\newcommand{\dG}{\mathtt G}

\newcommand{\tA}{\textrm A}
\newcommand{\tB}{\textrm B}
\newcommand{\A}{\mathcal A}
\newcommand{\B}{\mathcal B}
\newcommand{\C}{\mathcal C}
\newcommand{\D}{\mathcal D}
\newcommand{\E}{\mathcal E}
\newcommand{\F}{\mathcal F}
\newcommand{\G}{\mathcal G}
\newcommand{\M}{\mathcal M}
\newcommand{\HH}{\mathcal H}
\newcommand{\PP}{\mathcal P}

\newcommand{\Q}{\mathcal Q}
\newcommand{\Qb}{\bar{\mathcal Q}}
\newcommand{\Db}{{\bar{\Delta}}}

\newcommand{\pQ}{{\bf p}\mathcal Q}
\newcommand{\pQb}{{\bf p}\bar{\mathcal Q}}

\newcommand{\R}{\mathcal R}
\newcommand{\SSS}{\mathcal S}
\newcommand{\U}{\mathcal U}
\newcommand{\V}{\mathcal V}
\newcommand{\Y}{\mathcal Y}
\newcommand{\Z}{\mathcal Z}

\newcommand{\Pg}{{{\mathcal P}_{\rm gram}}}
\newcommand{\Pgint}{{{\mathcal P}^\circ_{\rm gram}}}
\newcommand{\Pgrc}{{{\mathcal P}_{\rm GRC}}}
\newcommand{\Pgrcint}{{{\mathcal P}^\circ_{\rm GRC}}}
\newcommand{\Pint}{{{\mathcal P}^\circ}}
\newcommand{\Ag}{{\bf A}_{\rm gram}}

\newcommand{\CC}{\mathbb C} 
\newcommand{\RR}{\mathbb R}
\newcommand{\ZZ}{\mathbb Z}
\newcommand{\FF}{\mathbb F}
\newcommand{\KK}{\mathbb K}

\newcommand{\Fnd}{\FF_q^{n^{\otimes d}}}
\newcommand{\Knd}{\KK^{n^{\otimes d}}}

\newcommand{\ceiling}[1]{\left\lceil{#1}\right\rceil}
\newcommand{\floor}[1]{\left\lfloor{#1}\right\rfloor}
\newcommand{\bbracket}[1]{\left\llbracket{#1}\right\rrbracket}

\newcommand{\inprod}[1]{\left\langle{#1}\right \rangle}


\newcommand{\beas}{\begin{eqnarray*}} 
\newcommand{\eeas}{\end{eqnarray*}} 

\newcommand{\bm}[1]{{\mbox{\boldmath $#1$}}} 

\newcommand{\sizeof}[1]{\left\lvert{#1}\right\rvert}
\newcommand{\wt}{{\rm wt}} 
\newcommand{\supp}{{\rm supp}} 
\newcommand{\dg}{d_{\rm gram}} 
\newcommand{\da}{d_{\rm asym}} 
\newcommand{\dist}{{\rm dist}} 
\newcommand{\ssyn}{s_{\rm syn}}
\newcommand{\sseq}{s_{\rm seq}}
\newcommand{\nullplus}{{\rm Null}_{>\vzero}}

\newcommand{\tworow}[2]{\genfrac{}{}{0pt}{}{#1}{#2}}
\newcommand{\qbinom}[2]{\left[ {#1}\atop{#2}\right]_q}

\newcommand{\Lovasz}{Lov\'{a}sz }
\newcommand{\etal}{\emph{et al.}}

\newcommand{\todo}{{\color{red} (TODO) }}

\title{DNA-Based Storage: Trends and Methods}

\author{S. M. Hossein Tabatabaei Yazdi\textsuperscript{1}, Han Mao Kiah\textsuperscript{2}, Eva Ruiz Garcia\textsuperscript{3}, Jian Ma\textsuperscript{4}, Huimin Zhao\textsuperscript{3}, Olgica Milenkovic\textsuperscript{1} \\
\thanks{This work was supported in part by the NSF STC Class 2010 CCF 0939370 grant and the Strategic Research Initiative (SRI) Grant conferred by the University of Illinois, Urbana-Champaign.}
\textsuperscript{1}Department of Electrical and Computer Engineering, University of Illinois, Urbana-Champaign\\
\textsuperscript{2}School of Physical and Mathematical Sciences, Nanyang Technological University, Singapore\\
\textsuperscript{3}Department of Bioengineering and Carl R. Woese Institute for Genomic Biology, University of Illinois, Urbana-Champaign\\
\textsuperscript{4}Department of Chemistry and Carl R. Woese Institute for Genomic Biology, University of Illinois, Urbana-Champaign}

\date{\today}

\maketitle

\begin{abstract}
We provide an overview of current approaches to DNA-based storage system design and accompanying synthesis, sequencing and editing methods. We also introduce and analyze a suite of new constrained coding schemes for both archival and random access DNA storage channels.
The mathematical basis of our work is the construction and design of sequences over discrete alphabets that avoid pre-specified address patterns, have balanced base content, and exhibit other relevant substring constraints. These schemes adapt the stored signals to the DNA medium and thereby reduce the inherent error-rate of the system.  
\end{abstract}
\vspace{-0.15in}
\section{Introduction}

Despite the many advances in traditional data recording techniques, the surge of Big Data platforms and energy conservation issues have imposed new challenges to the storage community in terms of identifying extremely high volume, non-volatile and durable  recording media. The potential for using macromolecules for ultra-dense storage was recognized as early as in the 1960s, when the celebrated physicists Richard Feynman outlined his vision for nanotechnology in the talk ``There is plenty of room at the bottom.'' Among known macromolecules, DNA is unique in so far that it lends itself to implementations of non-volatile recoding media of outstanding integrity (one can still recover the DNA of species extinct for more than 10,000 years) and extremely high storage capacity (a human cell, with a mass of roughly 3 pgrams, hosts DNA encoding 6.4 GB of information). Building upon the rapid growth of biotechnology systems for DNA synthesis and sequencing, two laboratories recently outlined architectures for archival DNA based storage in~\cite{church2012next,goldman2013towards}. The first architecture achieved a density of 700 TB/gram, while the second approach raised the density to 2 PB/gram. The success of the later method was largely attributed to the use of three elementary coding schemes, Huffman coding (a fixed-to-variable length entropy coding/compression method), differential coding (encoding the differences of consecutive symbols or the difference between a sequence and a given template) and single parity-check coding (encoding of a single symbol indicating the parity of the string). More recent work~\cite{grass2015robust} extended the coding approach used in~\cite{goldman2013towards} in so far by replacing single parity-check codes by Reed-Solomon codes~\cite{reed1960polynomial}. 

\begin{figure}
\begin{centering}
\includegraphics[scale=0.53]{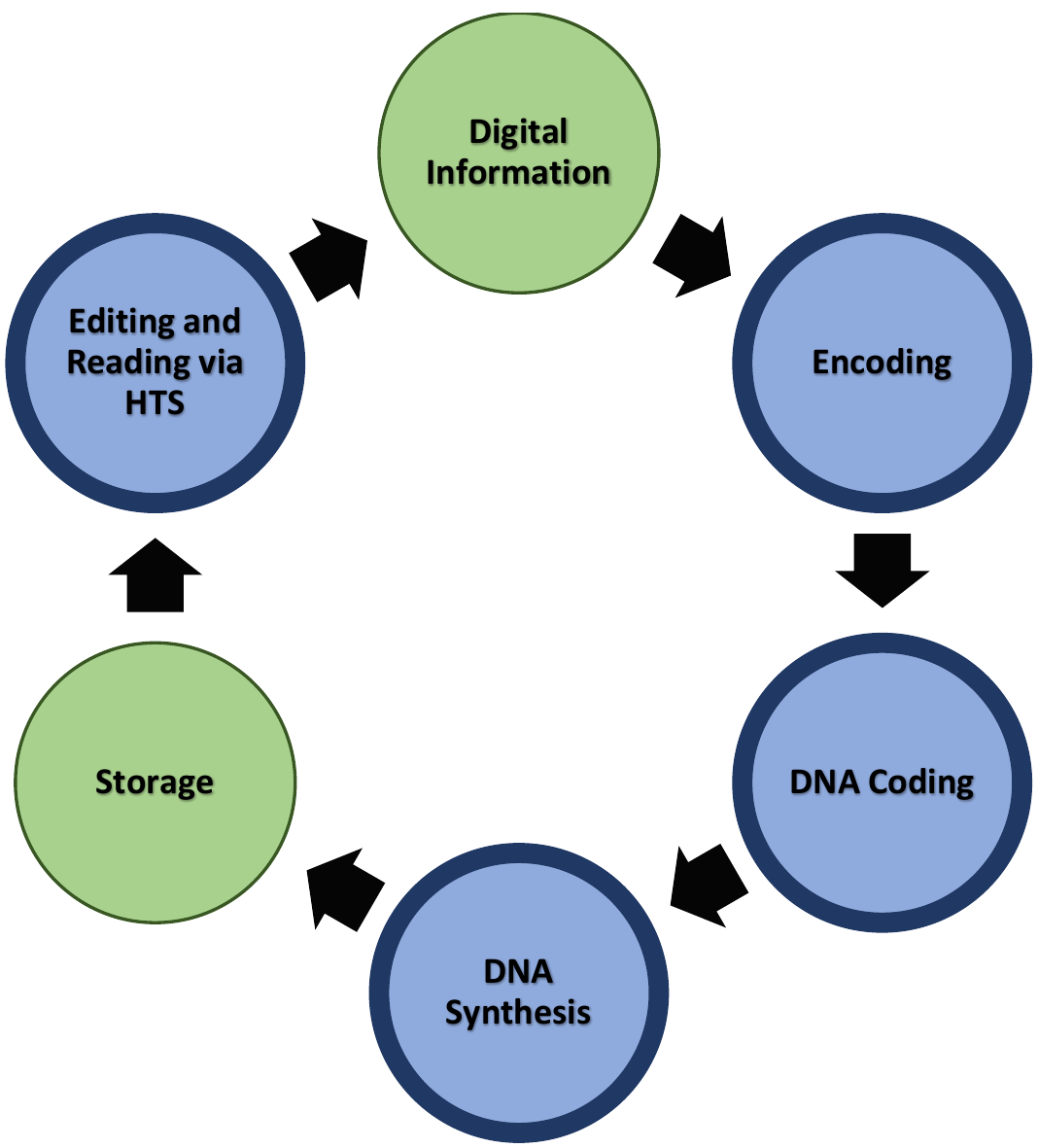} \label{fig:outline}

\par\end{centering}
\protect\caption{Block Diagram of Prototypical DNA-Based Storage Systems. A classical information source is encoded (converted into ASCII or some specialized word format, potentially compressed, and represented over a four letter alphabet); subsequently, the strings over four-letter alphabets are encoded using standard and DNA-adapted constrained and/or error-control coding schemes. The DNA codewords are synthesized, with potential undesired mutations (errors) added in the process, and stored. When possible, rewriting is performed via classical DNA editing methods used in synthetic biology. Sequencing is performed either through Sanger sequencing~\cite{schuster2008next}, if short information blocks are accessed, or via High Throughput Sequencing (HTS) techniques, if large portions of the archive are selected for readout.}
\vspace{-0.13in}
\end{figure}

All the aforementioned approaches have a number of drawbacks, including the lack of partial access to data -- i.e., one has to reconstruct the whole sequence in order to read even one base -- and the unavailability of rewrite mechanisms. Moving from a read only to a random access, rewritable memory requires a major paradigm shift in the implementation of the DNA storage system, as one has to append unique addresses to constituent storage DNA blocks that will not lead to erroneous cross-hybridization with the information encoded in the blocks; avoid using overlapping DNA blocks for increased coverage and subsequent synthesis, as they prevent efficient rewriting; ensure low synthesis (write) and sequencing (read) error rates of the DNA blocks. To overcome these and other issues, the authors recently proposed a (hybrid) DNA rewritable storage architecture with random access capabilities~\cite{yazdi2015rewritable}. The new DNA-based storage scheme encompasses a number of coding features, including constrained coding, ensuring that DNA patterns prone to sequencing errors are avoided; prefix synchronized coding, ensuring that blocks of DNA may be accurately accessed without perturbing other blocks in the DNA pool; and low-density parity-check (LDPC) coding for classically stored redundancy combating rewrite errors~\cite{gallager1962low}.
	
The shared features of current DNA-based storage architectures are depicted in Figure~\ref{fig:outline}. The green circles denote the source and media, while the blue circles denote processing methods applied on the source and media. The processes of Encoding and DNA Encoding add controlled redundancy into the original source of digital information or into the DNA blocks, respectively. This redundancy can be use to combat synthesis (write) and sequencing (access and read) errors~\cite{kiah2015codes, kiah2014codes, gabrys2015asymmetric}. Synthesis is the biochemical process of creating physical double-stranded DNA strings that reliably represent the encoded data strings. Synthesis thereby also creates the storage media itself -- the DNA blocks. Storage refers to some means of storing the DNA strings, and it represents a communication channel that transfers information from one point in time to another. In rewritable architectures~\cite{yazdi2015rewritable}, the Editing module refers to the process of creating mutations in the stored DNA strings (by deleting one or multiple substrings and potentially inserting other strings), while the Reading module refers to DNA sequencing that retrieves the content of selected DNA storage blocks and subsequent decoding.

In order to understand how errors occur during the read and write process, we start our exposition by describing state-of-the art synthesis, sequencing and editing methods (Sections~\ref{sec:synthesis},~\ref{sec:editing},~\ref{sec:sequencing}). We then proceed to discuss how synthesis, sequencing and editing methods are used in various DNA-storage paradigms (Section~\ref{sec:architectures}), and the accompanying coding techniques identified with different types of synthesis and sequencing errors. New constrained coding techniques for rewritable and random access systems, and their relationship to classical codes for magnetic and optical storage, are described in Section~\ref{sec:coding}.

Given the semi-tutorial and interdisciplinary nature of this manuscript, we refer readers with a limited background in synthetic biology to Appendix~\ref{sec:appendix} for a glossary of terms used throughout the paper.

\section{DNA Sequence Synthesis} \label{sec:synthesis}

De novo DNA synthesis is a powerful biotechnology that enables the creation of DNA sequences without pre-existing templates. Synthesis tools have a myriad of applications in different research areas, ranging from traditional molecular biology to emerging fields of synthetic biology, nanotechnology and data storage.
Vaguely speaking, most technologies for large-scale DNA synthesis rely on the assembly of pools of oligonucleotide building blocks into increasingly larger DNA fragments. The current high cost and small throughput of de novo synthesis of these building blocks represents the main limitation for widespread implementations of DNA synthesis systems: as an example, oligo synthesis methods via \emph{phosphoramidite column-based synthesis}, described in subsequent sections, may cost as much as ~\$0.15 per nucleotide~\cite{kosuri2014large}. The maximum length of the produced oligostrings lies in the range 100-200 nts~\cite{kosuri2014large}. Hence, the synthesis of long DNA oligos using dozens of building blocks can cost anywhere from hundreds to thousands of US dollars. Therefore, it is imperative to develop new, high-quality, robust, and scalable DNA synthesis technologies that offer synthetic DNA at significantly more affordable prices. This is in particular the case for massive DNA-based storage systems, which may potentially require billions of nucleotides.  

Among the most promising synthesis technologies is the so called \emph{microarray-based} synthesis methods; more than ten-to-hundreds of thousands oligos can be synthesized per one microarray, in conjunction with a decrease in the reagent consumption. For large scale DNA synthesis projects, the price of microarray-based synthesis is roughly ~\$0.001 per nucleotide~\cite{kosuri2014large,tian2009advancing}. Similarly to the case of phosphoramidite column-based synthesis, the length of microarray synthesized oligos usually does not exceed $200$ nt. However, oligos synthesized in microarrays typically suffer from higher error rates than those generated by phosphoramidite column methods. Nevertheless, microarrays are the preferred synthesis tool for generating customized DNA-chips or for performing gene synthesis. Many projects are underway to bridge the gap between these two extremes, hight-cost and high-accuracy and low-cost, low-accuracy strategies and hence reduce the limitations of the corresponding methods~\cite{ma2012dna,ma2012error}. 

To provide a better understanding of the basic principles of DNA-based storage and the limitations that need to be overcome in the writing process, we first describe different DNA synthesis methods from nucleotides to larger DNA molecules. We then discuss recent techniques that aim to improve the quality and reliability of the synthesized sequences.

\subsection{Chemical Oligonucleotide Synthesis}

Chemical synthesis of single stranded DNA originated more than $60$ years ago, and since the 1950's, when the first oligonucleotides were synthesized~\cite{michelson1955nucleotides,hall1957644,gilham1958studies}, four different chemical methods have been developed. These methods are named after the major reagents included in the process, and include i) H-phosphonate; ii) phosphodiester; iii) phosphotriester; and iv) phosphite triester/phosphoramidite. A detailed description of these methods may be found in~\cite{roy2013synthesis,reese2005oligo}, and here we only briefly review the advantages and disadvantages of these methods.

The H-phosphonate method was first described in~\cite{hall1957644}, and it derives its name from the use of H-phosphonates nucleotides as building blocks. This approach was later refined in~\cite{froehler1986synthesis,garegg1986nucleoside}, where the H-phosphonate chemistry was improved to synthesize deoxyoligonucleotides on a solid support by using different oligo coupling (stitching) agents that expedite the reactions. The phosphodiester method was introduced in~\cite{gilham1958studies,khorana1957syntheses}. The main contribution of the method was the production of \emph{protected dinucleotide monoposhpates}, which prevented undesired elongation. 
Unfortunately, the approach also had one major drawback -- the linkages between nucleotides were unprotected during the elongation step of the oligonucleotide chain, which allowed for the creation of branched oligonucleotides.
The phosphotriester approach was also first published in the 50s~\cite{michelson1955nucleotides} and later improved by Letsinger~\cite{letsinger1965oligonucleotide,letsinger1969nucleotide} and Reese~\cite{reese1968oligonucleotide} using different reagents to protect the phosphate group in the internucleotide linkages. This approach also prevented the formation of branched oligonucleotides. Nevertheless, all the previously described methods and variants thereof proved to be inefficient and time consuming. 

In the mid-seventies, a major advantage in synthesis technology was reported by Letsinger~\cite{letsinger1976synthesis}, solving in part a number of problems associated with other existing methods. His method was termed the \emph{phosphite triester approach}. The basic idea behind the approach was that the reagent phosphorochloridite reacts with nucleotides faster than its chloridate counterpart used in previous approaches. In addition to expediting their underlying reactions, bifunctional phosphorodichloridites unfortunately also produced undesirable side products such as symmetric dimers.
A modified method that precluded the drawback of side products was developed by Caruthers \etal~\cite{beaucage1981deoxynucleoside}. The authors of ~\cite{beaucage1981deoxynucleoside} used a different type of nucleoside phosphites that were more stable, reacted faster, and produced higher yields of the desired dinucleoside phosphite. The resulting method was named \emph{phosphoramidite synthesis}. Another important contribution includes the technology described in~\cite{sinha1984polymer}, where the use of stable and easy-to-prepare phosphoramidites facilitated the automation of oligo synthesis in solid-phase, making it the method-of-choice for chemical synthesis.

\subsection{Oligo Synthesis Platforms}

\begin{figure*}
\begin{centering}
\includegraphics[scale=0.46]{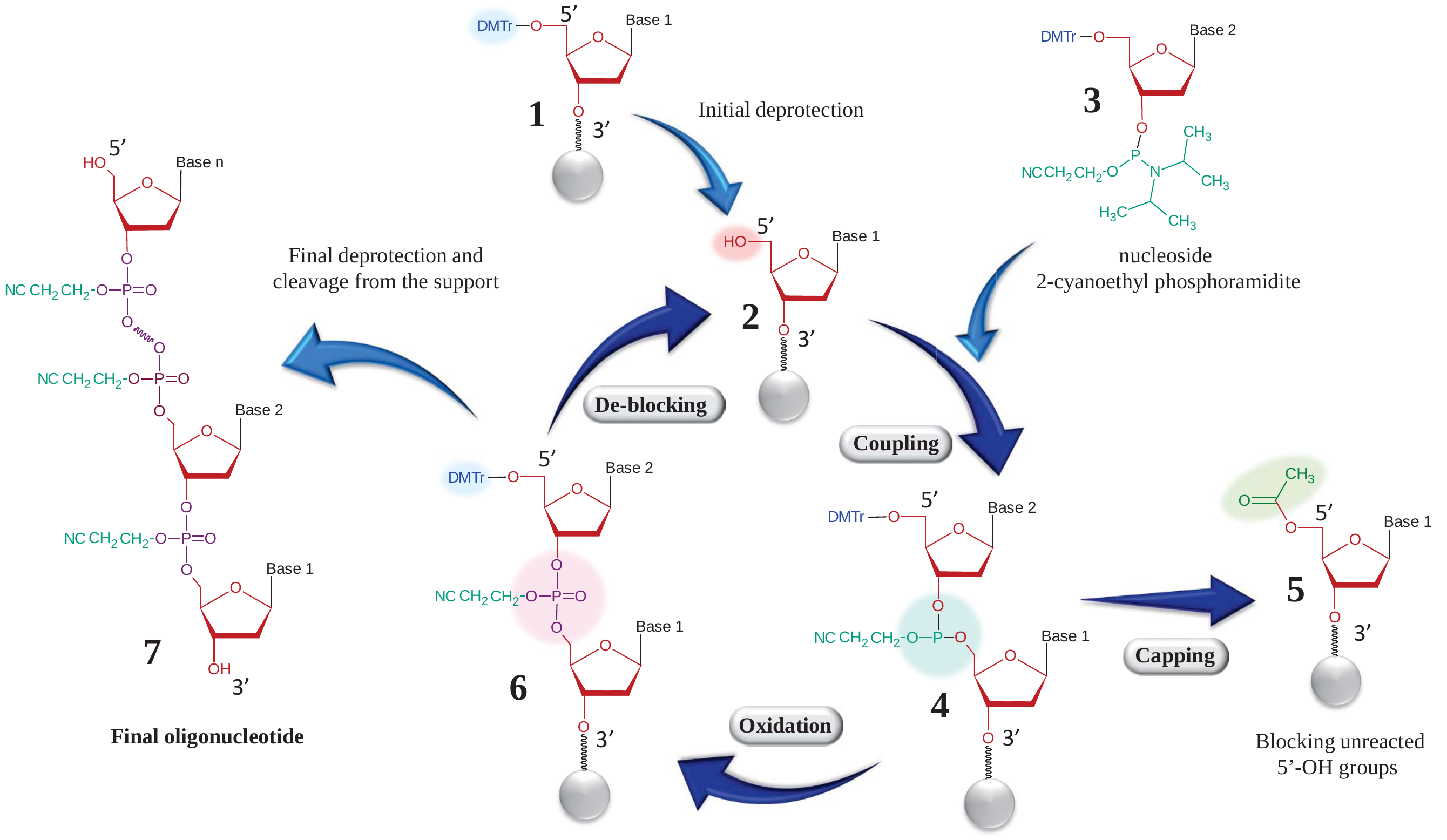} \label{fig:column}

\par\end{centering}
\protect\caption{Main Steps of Column-Based Oligo Synthesis process of Section~\ref{sec:column}. The first step in DNA synthesis cycle is the deprotection of the support-bound nucleoside at the 5' terminal end (1, highlighted in blue) by removal of the DMTrgroup. This step lead a nucleoside with a 5' OH group (2, highlighted in red). During the coupling step an activated nucleoside (3) react with the 5' OH group of the support-bound nucleoside (2) generating a dinucleotide phosphoramidite (4) (formation of phosphitetriester, highlighted in green-blue). In the capping step, unreacted 5' OH are blocked by acetylation (5, highlighted in green) to prevent further chain extension. In the last step of the cycle the unstable phosphitetriester (in green-blue) is oxidized to phosphate linkage (6, highlighted in purple) which is more stable in the chemical conditions of the following synthesis steps. The cycle is repeated for each nucleoside addition. After the last step of the synthesis of the entire oligonucleotide, the final product needs to be cleavage from the solid support and deprotect the 5' terminal end. In red is the pentose of the nucleoside, in blue the dimethoxytrityl (DMTr) protecting group, in green-blue the 2-cyanoethyl phosphoramiditegroup, and in purple the phosphate group. Grey spheres represent the solid support in which the growing oligo is attached. Circles highlight the group that is modified in each step.}
\end{figure*}

\subsubsection{Column-Based Oligo Synthesis} \label{sec:column}

The standard phosphoramidite oligonucleotide synthesis operates via stepwise addition of nucleotides to the growing chain which is immobilized on a solid support (Figure~\ref{fig:column}). Each addition cycle consists of four chemical steps: i) de-blocking; ii) coupling or condensation; iii) capping; and iv) oxidation~\cite{roy2013synthesis}. At the beginning of the synthesis process, the first nucleotide, which is attached to a solid substrate, is completely protected at all of its active sites. Therefore, to make a reaction possible and include a second nucleotide, it is necessary to remove the dimethoxytrityl (DMT) protecting group from the 5'-end by addition of an acid solution. The removal of the DMT group generates a reactive 5'-OH group (De-blocking step). Subsequently, a coupling step is performed via condensation of a newly activated DMT-protected nucleotide and the unprotected 5'-OH group of the substrate-bound growing oligostrand through the formation of a phosphite triester link (Coupling or Condensation step). After the coupling step, some unprotected 5'-OH groups may still exist and react in later stages of additions of nucleotides leading to oligos with \emph{deletion and bursty deletion errors}. To mitigate this problem, a capping reaction is performed by acetylation of the unreactive nucleotides (Capping step). Finally, the unstable phosphite triester linkage is oxidized to a more stable phosphate linkage using an iodine solution (Oxidation step). The cycle is repeated iteratively to obtain an oligonucleotide of the desired sequence composition. At the end of the synthesis, the oligonucleotide sequence is deprotected, and cleaved from the support to obtain a completely functional unit.

\subsubsection{Array-Based Oligo Synthesis}

In the 90s, Affymetrix developed a method for chemical synthesis of different polymers combining photolabile protecting groups and photolithography~\cite{fodor1991light,pease1994light}. The Affymetrix solution uses a photolithographic mask to direct UV light in a targeted manner, so as to selectively deprotect and activate 5' hydroxyl groups of nucleotides that should react with the nucleotide to be incorporated in the next step. The mask is designed to expose specific sites on the microarray to which new nucleotides will be added, with others sites being masked. Once synthesis is completed, the oligos are released from the array support and recovered as a complex mixture (pool) of sequences.

A number of other, related methods have been developed for the purpose of synthesizing oligostrands on microarrays~\cite{gao2004situ}. For instance, the method developed by Agilent uses Ink-jet-based printing, where with high precision, picoliters of each incorporated nucleotide and activator can be spotted (deposited) at specific sites on an array. This ink-jet method mitigates the need for using photolithography masks~\cite{hughes2001expression}. In an alternative method commercialized by NimbleGen Systems, the photolithography masks are superseded by a virtual mask that is combined with digital programmable mirrors to activate specific locations on the array~\cite{singh1999maskless,nuwaysir2002gene}. CustomArray (former CombiMatrix) developed a technology in which thousands of microelectrodes control acid production by an electrochemical reaction to deprotect the growing oligo at a desired spot~\cite{ghindilis2007combimatrix}. In addition, oligo synthesis is implemented within a multi-chamber microfluidic device coupled to a digital optical device that uses light to produce acid in the chambers~\cite{kong2007parallel}. Masking and printing errors may introduce both \emph{substitution} and \emph{insertion and deletion} errors, and when multiple sequences are synthesized simultaneously, the error patterns within different sequences may be correlated, depending on the location of their synthesis spots.
 
Both solid-phase and microarray technologies exhibit a number of challenges that need to be overcome to reduce error rates and increase throughout. Side reactions such as depurination~\cite{efcavitch1985depurination,leproust2010synthesis} and reaction inefficiencies during the stepwise addition of nucleotides~\cite{roy2013synthesis,reese2005oligo} reduce the desired yield, and generate errors in the sequence especially when synthesizing long oligostrands. In particular, these processing problems introduce both \emph{substitution} and \emph{insertion and deletion} errors. Thus, a purification step is usually necessary to identify and discard undesirable erroneous sequences. High-performance liquid chromatography and polyacrylamide gel electrophoresis can be used to eliminate truncated products, but both are expensive and time-consuming, and single insertions and deletions or substitution errors in the sequence often cannot be removed. Nevertheless, by optimizing chemical reaction and conditions the fidelity can be increased~\cite{leproust2010synthesis}.

\subsubsection{Complex Strand and Gene Synthesis}

Traditionally, to generate DNA fragments of length several hundred nucleotides, a set of shorter length oligostrands is fused together by either using ligation-based or polymerase-based reactions. Ligation-based approaches usually rely on thermostable DNA ligases that ligate phosphorylated overlapping oligos in high stringency conditions~\cite{au1998gene}. In polymerase-based approaches (Polymerase cycling assembly - PCA) oligos with overlapping regions are used to generate progressively longer double-stranded sequences~\cite{stemmer1995single}. After assembly, synthesized sequences need to be PCR amplified, cloned, and verified, thus increasing the cost of production.
Another approach developed by Gibson \etal~\cite{gibson2009synthesis} exploits yeast in vivo recombination to assemble a set of more than $30$ oligos together with a plasmid, all in one step. The same group also synthesized the mouse mitochondrial genome from $600$ overlapping oligos using an isothermal assembly method~\cite{gibson2010chemical}.

Although microarray synthesis reduces the price of oligonucleotides, there are two major challenges that still hamper its use. First, hundreds of thousands of oligonucleotides can be made on a single microarray, but each oligo is produced in very small amounts. Second, the oligostrands are cleaved from the array all at once as a large heterogeneous pool that subsequently leads to difficulties in sequence assembly and cross-hybridization. A number of strategies have been recently developed to solve these problems. For example, PCR amplification increases the concentration of the oligos before assembly that combined with hybridization selection reduces the incorporation of oligonucleotides containing undesirable synthesis errors~\cite{tian2004accurate}. A modification of this approach, based on hybridization selection embedded in the assembly process and coupled with the optimization of oligo design and assembly conditions was reported in~\cite{borovkov2010high}. Still, large pools of oligos (>$10000$) increase difficulties in sequence assembly. Two different strategies have been described where subpools of oligos involved in a particular assembly were isolated, thus partially avoiding cross-hybridization. Kosuri \etal~\cite{kosuri2010scalable} used predesigned barcodes to amplify subpools of oligos, and in a second step removed the barcodes by digestion. In another approach, the microarray was physically divided in sub-arrays that enabled performing amplification and assembly separately in each microwell~\cite{quan2011parallel}.

\begin{figure}
\begin{centering}
\includegraphics[scale=0.45]{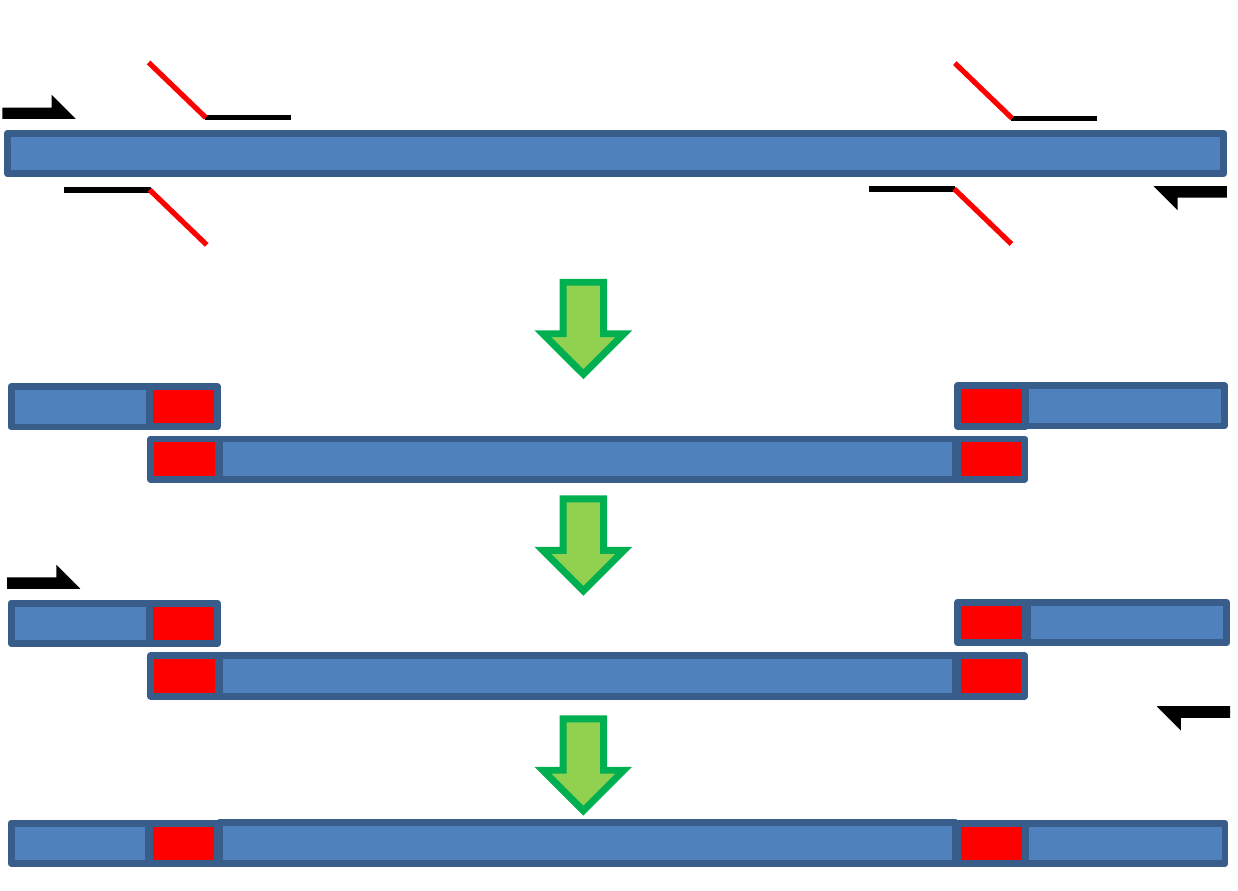} \label{fig:gblock}

\par\end{centering}
\protect\caption{Rewriting (Deletion and Insertion Edits) via gBlocks. This method is used when edits of relatively short length are required, as it is cost efficient and simple. Primers corresponding to unique contexts in the encoded DNA are used to access the edit region, which is subsequently cleaved and replaced by the gBlock.}
\end{figure}

\begin{figure*}
\begin{centering}
\includegraphics[scale=0.5]{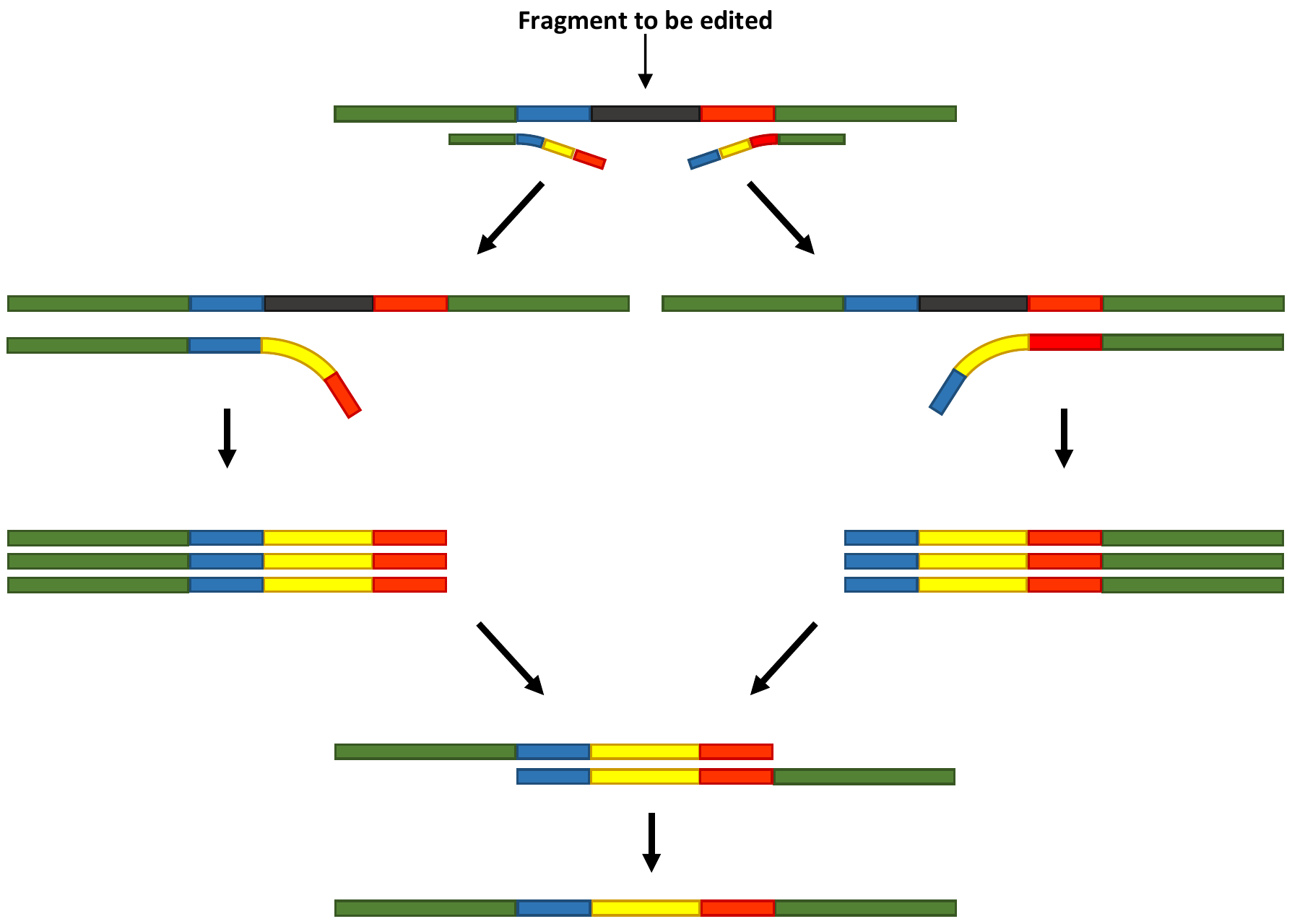} \label{fig:oepcr}

\par\end{centering}
\protect\caption{Rewriting (Deletion and Insertion Edits) via OEPCR. OEPCR allows for incorporating customized sequence changes via primers used in amplification reactions. As the primers have terminal complementarity, two separate DNA fragments may be amplified and fused into a single sequence without using restriction endonuclease sites. Overlapping fragments are fused together in an extension reaction and PCR amplified.}
\end{figure*}

\subsubsection{Error Correction}

Despite having elaborate biochemical error removal processes in place, some residual errors tend to remain in the synthesized pool and additional errors arise during the assembly phase. A number of error-correction strategies have been reported in the literature~\cite{kosuri2014large,ma2012dna,ma2012error}. Many of the current error-removal techniques rely on DNA mismatch recognition proteins. Denaturation and re-hybridization steps lead to double-stranded DNA with mismatches between erroneous bases and the corresponding correct bases. The disrupted sites are recognized and/or cleaved by mismatch recognition proteins. MutS is a protein that binds unpaired bases and small DNA loops (i.e., small unmatched substrings in DNA that protrude from the double helix). After denaturation and re-hybridization, MutS detects and binds the mismatched regions that are later removed by gel electrophoresis. This strategy reduces the error-rate to $1$ nucleotide per $10$ Kb~\cite{carr2004protein}. ``Consensus shuffling'' is a variation of the MutS method where mismatch-containing pieces are captured by column-immobilized MutS proteins, and error-free fragments are eluted~\cite{binkowski2005correcting}. In other variations of this method, two homologs of MutS immobilized in cellulose columns can reduce the error rate to $0.6$ nucleotides per Kb at a very low cost~\cite{wan2014error}. On the other hand, in the MutHLS approach, MutS binds unpaired bases, while the protein MutL links the MutH endonuclease to the MutS bound sites that cleave the erroneous heteroduplexes. The correct sequences are recovered by gel electrophoresis~\cite{smith1997removal}. Similarly, resolvases ~\cite{fuhrmann2005removal} and single-strand nucleases~\cite{till2004mismatch,oleykowski1998mutation} may also be used to recognize and cleave mismatched sites in DNA heteroduplexes. It is worth pointing out that CEL endonuclease, its commercial version Surveyor\textsuperscript{TM} nuclease (Transgenomic, Inc.) or a commercial CEL-based enzymatic cocktail, ErrASE, that recognizes and nicks at the base-substitution mismatch, is commonly used in practice due to its broad substrate specificity; it can reduce the error rate up to $1$ nucleotide per $9.6$ Kb~\cite{saaem2011error,dormitzer2013synthetic}. 

The introduction of Next Generation Sequencing (NGS) platforms as high throughput purification methods opened new possibilities for error-free DNA synthesis. Matzas \etal~\cite{matzas2010high} combined a next-generation pyrosequencing platform with a robotic system to image and pick beads containing sequence-verified oligonucleotides. The estimated error rate using this approach is $1$ nucleotide error per $21$ Kb. One limitation of this method is that the ``pick-and-place'' recovery system is not accurate enough, due to the small size of clonal beads, to satisfy the increasing demand for long length DNA strands (involving $104$ building blocks)~\cite{lee2015high}. A new NGS-based method was recently announced, where specific barcoded primers were used to amplify only those oligos with the correct sequence~\cite{kim2012shotgun,schwartz2012accurate}. Similarly, a new method termed ``sniper cloning'' has been reported in~\cite{lee2015high}. There, NGS platform beads containing sequence-verified oligonucleotides are recovered by ``shooting'' a laser pulse. This laser technology enables cost-effective, high throughput selective separation of correct fragments without cross-contamination.

As a parting note, we observe that even single substitution errors in the synthesis process may be detrimental for applications in biological and medical research. This is \emph{not the case} for DNA-based storage systems, where the DNA strands are used as storage media which may have a non-negligible error rate. Synthesis errors may be easily combated through the introduction of carefully designed parity-checks of the information strings, as will be discussed in subsequent sections.

\section{DNA Editing} \label{sec:editing}

Once desired information is stored in DNA by synthesizing properly encoded heteroduplexes, it may be rewritten using classical \emph{DNA editing} techniques. DNA editing is the process of adding very specific point mutations (often with the precision of a few nucleotides) or deleting and inserting DNA substrings at tightly controlled locations. In the latter case, one needs to synthesize readily usable short-to-medium length DNA fragments. For this purpose, two techniques are commonly used: gBlocks Gene Fragments~\cite{idtdna} (see Integrated DNA Techologies) as building blocks for insertion and deletion edits, and Overlap-Extension PCR (OEPCR)~\cite{higuchi1988general} as a means of adding the mutated blocks.

gBlocks are double-stranded, precisely content-controlled DNA blocks that may be used for applications as diverse as gene construction, PCR and qPCR control, recombinant antibody research, protein engineering, CRISPR-mediated genome editing and general medical research~\cite{jansen2002identification}. They are usually constructed at very low cost (fraction of a dollar) using \emph{gene fragments libraries}, i.e., pools of short DNA strings that contain up to $18$ consecutive bases  of type N (any nucleotide) or K (Keto). The libraries and library products are carefully tested for correct length via capillary electrophoresis, sequence composition via mass spectrometry; consensus protocols are used in the final verification stage to reduce any potential errors. The last stage, and additional quality control testing ensures that at least 80\% of the generated pool contains the desired string. For strings with complex secondary structure, this percentage may be significantly lower. This calls for controlling the secondary structure of the products whenever the applications allows for it. Such is the case for DNA-based storage, and methods for designing DNA codewords with no secondary structure (predicted to the best extent possible via combinatorial techniques) were described in~\cite{milenkovic2006design}.

DNA substring editing is frequently performed via specialized PCR reactions. Of particular use in DNA rewriting is the process of OEPCR, illustrated in Figure~\ref{fig:oepcr}. IN OEPCR, one uses two primers to flank two ends of the string to be edited. For fragment deletion (splicing), the flanking primers act like zippers that need to join over the segment to be sliced. Furthermore, the primer at the end to be joined is designed so that it has an overhanging part complementary to the overhanging part of the other primer. Via controlled hybridization, the DNA strands are augmented by a DNA insert that is also complementary to the underlying DNA strand. Upon completion of this extension, classical PCR amplification is performed for the elongated sequence primers and the inserted overlapping fragments of the sequences are fused. Note that this method does not requiring restriction sites or enzymes. OEPCR is mostly used to insert oligonucleotides of lengths longer than $100$ nucleotides. In OEPCR the sequence being modified is used to make two modified strands with the mutation at opposite ends, using the method outlined above. After denaturation, the strands are mixed, leading to different hybridization products. Of all the products, only one will allow for polymerase extension via the introduction of a primer -- the heterodimer without overlap at the 5' end. The duplex created by the polymerase is denatured once again and another primer is hybridized to the created DNA strand, introducing a sequence contained in the first primer. DNA replication consequently results in an extended sequence containing the desired insert.

\begin{figure*}
\begin{centering}
\includegraphics[scale=0.65]{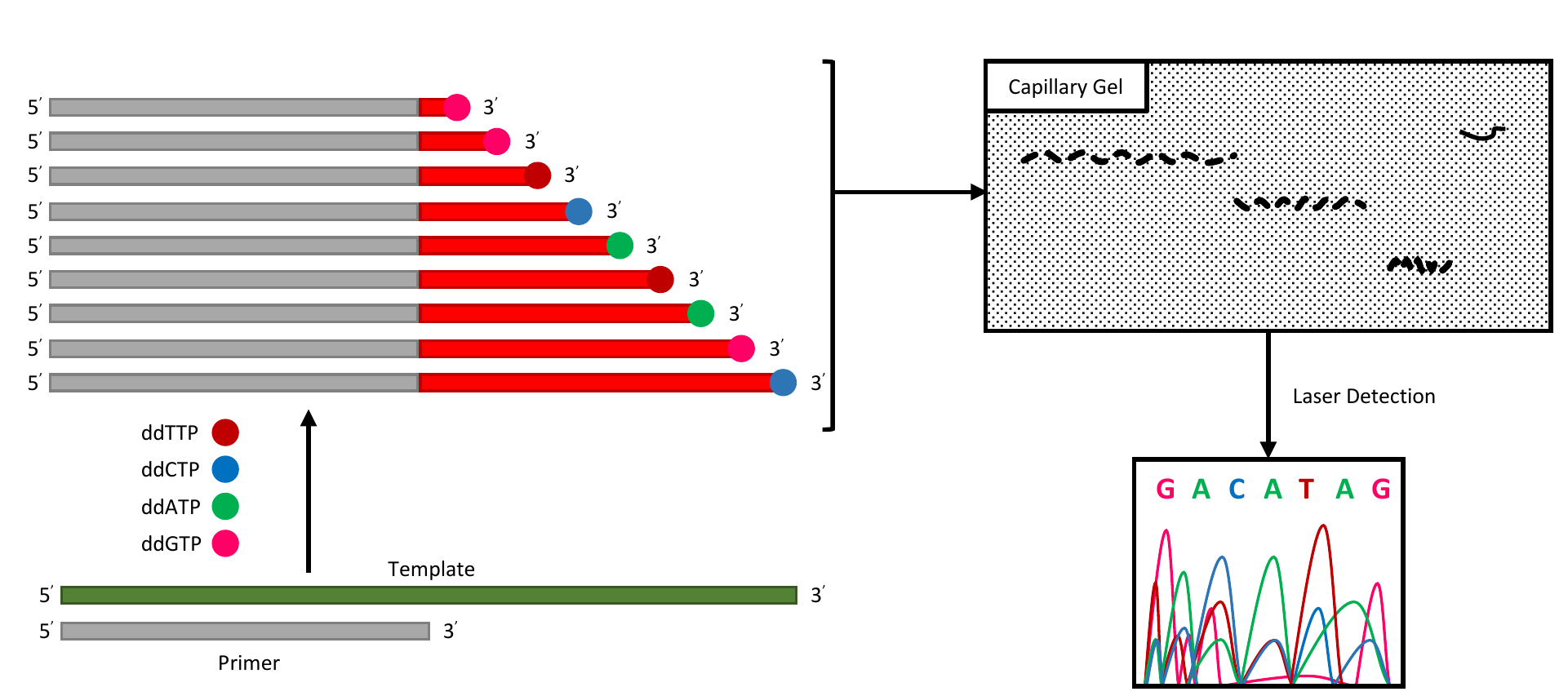} \label{fig:sanger}

\par\end{centering}
\protect\caption{Main steps of the Sanger sequencing protocol. In the first step, a pool of DNA fragments is sequenced via synthesis. Synthesis terminates whenever chemically inactive versions of nucleotides (dd*TP) are incorporated into the growing chains. These inactive nucleotides are fluorescently labeled to uniquely determine their bases. In the second step, the fragments are sorted by length using capillary gel methods. The terminal step involves reading the last bases in the fragments using laser systems.}
\vspace{-0.13in}
\end{figure*}

\vspace{-0.05in}

\section{DNA Sequencing} \label{sec:sequencing}

The goal of DNA sequencing is to read the DNA content, i.e., to determine the exact nucleotides and their order in a DNA molecule. Such information is critical in understanding both basic biology and human diseases as well as for developing nature-inspired computational platforms. 

Sanger \etal~\cite{Sanger1977} first developed sequencing methods to sequence DNA based on chain termination (see Figure~\ref{fig:sanger} for an illustration). 
This technique, which is commonly referred to as Sanger sequencing, has been widely used for several decades and it is still being used routinely in numerous laboratories.
The automated and parallelized approaches of Sanger sequencing directly led to the success of the Human Genome Project~\cite{Lander2001} and the genome sequencing projects of other important model organisms for biomedical research (e.g., mouse~\cite{Waterston2002}). 
The availability of these entire genomes has provided scientists with unprecedented opportunities to make novel discoveries for genome architecture and genome function, trajectory of genome evolution, and molecular bases of phenotypic variation and disease mechanisms. 

However, in the past decade, the development of faster, cheaper, and higher-throughput sequencing technologies has dramatically expanded the reach of genomic studies.
These ``next-generation sequencing'' (NGS) technologies, as opposed to Sanger sequencing which is considered as first-generation, have been one of the most disruptive modern technological advances.
In general, the NGS technologies have several major differences when compared to Sanger sequencing.
First, electrophoresis is no longer needed for reading the sequencing output (i.e., substring lengths) which is now typically detected directly. Second, more straightforward library preparations that do not use DNA clones have become a critical part of sequencing workflow.
Third, tremendously large number of sequencing reactions are generated in parallel with ultra-high throughput. 
A demonstration of the significant NGS technology development is the cost reduction.
Around the year 2001, the cost of sequencing a million base-pairs was about \$5,000;
but it only costs about \$0.05 in mid 2015 (http://www.genome.gov/sequencingcosts/).
In other words, it will cost less than \$5,000 to sequence an entire human genome with $30$x coverage.
This cost keeps dropping every few months due to new developments in sequencing technology. 
However, a clear shortcoming of NGS versus Sanger technologies has been data quality. 
The read lengths are much shorter and the error rate is higher as compared to Sanger sequencing. 
For instance, the read length from Illumina sequencing platforms ranges from $50$bps to $300$bps,
making subsequent genome assembly extremely difficult, especially for genomes with a large proportion of repetitive elements/substrings. The error-rates of latest Illumina sequencing platforms, such as HiSeq 2500 are less than $1\%$, and the errors are highly non-uniformly distributed along the sequenced reads: the terminal 20$\%$ of nucleotides have orders of magnitude higher error-rates than the remaining $80\%$ of initial bases. 

The first NGS platform was introduced by $454$ Life Sciences (acquired by Roche in $2007$).
Although Roche will shut down $454$ in $2016$, $454$ platforms have made significant contributions to both NGS technology development and biological applications, including the first full genome of a human individual using NGS~\cite{Wheeler2008}.
The $454$ platform utilizes pyrosequencing. 
Briefly, pyrosequencing operates as follows. DNA samples are first fragmented randomly. 
Then each fragment is attached to a bead and emulsion PCR is used to make each bead contain many copies of the initial fragment.
The sequencing machine contains numerous picoliter-volume wells, each containing a bead.
In pyrosequencing, luciferase is used to produce light, initiated by pyrophosphate when a nucleotide is incorporated at each cycle during sequencing.
One drawback of $454$ sequencing is that multiple incorporation events occur in homopolymers.
Therefore, as the length of a homopolymer is reflected by the light intensity, a number of sequencing errors arise in connection with homopolymers. We remark that such errors were accounted for in a number of DNA-storage implementations, even those using other sequencing platforms which typically do not introduce homopolymer errors.

The SOLiD planform, developed by Applied Biosystems (merged with Invitrogen to become Life Technologies in $2008$), was introduced in $2007$. SOLiD uses sequencing by ligation; 
i.e., unlike $454$, DNA ligase is used instead of polymerase to identify nucleotides.
During sequencing, a pool of possible oligonucleotides of a certain length are labeled according to the sequenced position. 
These oligonucleotides are ligated by DNA ligase for matching sequences. 
Before sequencing, the DNA is amplified using emulsion PCR.
Each of the resulting beads contains single copies of the same DNA molecule.
The output of SOLiD is in color space format, an encoded form of the nucleotide sequences with four colors representing $16$ combinations of two adjacent bases.

The most frequently used sequencing platform so far has been Illumina. 
It's sequencing technology was developed by Solexa, which was acquired by Illumina in $2007$.
The method is mainly based on reversible dye-terminators that allow the identification of nucleotide bases when they are introduced into DNA strands.
DNA samples are first randomly fragmented and primers are ligated to both ends of the fragments.
They are then attached on the surface of the flow cell and amplified -- in a process also known under the name bridge amplification -- so that local clonal DNA colonies, called ``DNA clusters'', are created.
To determine each nucleotide base in the fragments, sequencing by synthesis is utilized.
A camera takes images of the fluorescently labeled nucleotides to enable base calling. 
Subsequently, the dye, along with the terminal 3' blocker, is removed from the DNA to allow for the next cycle to begin with multiple iterations.The most frequently encountered errors in Illumina data are simple substitution errors. Much less common are deletion and insertion errors, and there is an indication that sequencing error rates are higher in regions in which there are homopolymers exceeding lengths $15-20$~\cite{ross2013characterizing}. 
Substitution errors arise when nucleotides are incorporated at different positions in the fragments of a cluster during the same cycle. They are also caused by clusters from more than one DNA fragment, resulting in mixed signals during the base calling step.
Illumina sequencers have been used in numerous NGS applications, ranging from whole-genome sequencing, whole-exome sequencing, to RNA sequencing, ChIP sequencing and others. 
The Illumina HiSeq 2500 system can generate up to $2$ billion single-end reads (in $250$ bp) per flow cell with $8$ lanes. The recently announced HiSeq $4000$ system can produce up to $5$ billion single-end reads per flow cell.

In addition, several other types of sequencing technologies have been developed in recent years, with the Pacific Biosciences (PacBio) single-molecule real time (SMRT) technology and the Oxford Nanopore's nanopore sequencing systems being the most promising ones. 
In SMRT, no amplification is needed and the sequencer observes enzymatic reaction in real time. 
It is also sometimes referred as ``third-generation sequencing'' because it does not require any amplification prior to sequencing.
The most significant advantage of PacBio data is the much longer read length as compared to other NGS technologies. SMRT can achieve read lengths exceeding $10$ Kbases, making it more desirable for finishing genome assemblies. Another advantage is speed -- run times are much faster. However, the cost of PacBio sequencing is fairly high, amounting to a few dollars per million base-pairs. 
Furthermore, SMRT error rates are significantly higher than those of Illumina sequencers and the throughput is much lower as well.

Oxford Nanopore is considered another third-generation technology. Its approach is based on the readout from eletrical signals when a single-stranded DNA sequence passes through a nanoscale hole made from proteins or synthetic materials.
The DNA passing through the nanopore would change its ion current, allowing the sequencing process to recognize nucleotide bases. 
Oxford Nanopore has developed a hand-held device called MinION, which has been available to early users. MinION can generate more than $150$ million bases per run. 
However, the error rate is significantly higher than other technologies and it is still being improved. Some of the errors were identified in~\cite{gabrys2015asymmetric} as \emph{asymmetric errors}, caused by two bases creating highly similar current impulse responses.

Significant challenges of NGS still remain, in particular data analysis problems arising due to short read length. One major step after having the sequencing reads is to assemble reads into longer DNA fragments. 
Most of these assemblers follow a multi-stage procedure: correcting raw read errors, constructing contigs (i.e. contiguous sequences obtained via overlapping reads), resolving repeats, and connecting contigs into scaffolds using paired-end reads. Most de novo assemblers utilize the de Bruijn graph (DBG) data structure to represent large number of input short reads. EULER~\cite{Pevzner2001} pioneered the use of DBG in genome assembly.
In recent years, several NGS assemblers (such as Velvet~\cite{Zerbino2008}, ALLPATHS-LG~\cite{Gnerre2011}, SOAPdenovo~\cite{Li2010}, ABySS~\cite{Simpson2009}, SGA~\cite{Simpson2012}) have shown promising performances.

\section{Archival DNA-Based Storage} \label{sec:architectures}

\subsection{The Church-Gao-Kosuri Implementation}

The first large-scale archival DNA-based storage architecture was implemented and 
described in the seminal paper of Church \etal~\cite{church2012next}. In the proposed approach, 
user data was converted to a DNA sequence via a symbol-by symbol mapping, encoding each data 
bit $0$ into A or C, and each data bit $1$ into T or G. Which of the two bases is used for encoding
a particular bit is determined by a runlenghth constraint, i.e., one base is chosen randomly as long as it prohibits 
homopolymer runs of length greater than three. Furthermore, the choice of one of the two bases 
enables control of the GC content and secondary structure within the DNA data blocks.

To illustrate the feasibility of their approach, the authors of~\cite{church2012next} 
encoded in DNA a HTML file of size $5.27$ MB. The file included $53,426$ words, $11$ JPG images and one Java Script file.
In order to eliminate the need for long synthetic DNA strands that are hard
to assemble, the file was converted into $54,898$ blocks of length $159$ oligonucleotides. Each 
block contained $96$ information oligonucleotides, $19$ oligonucleatides for addressing, and $22$ oligonucleotides for a 
common sequence used for amplification and sequencing. The $19$ oligonucleotide addresses corresponded to binary encodings of consecutive integers, starting from $00\ldots001$. 

The oligonucleotide library was synthesized using Ink-jet printed, high-fidelity
DNA microchips~\cite{leproust2010synthesis}, described in Section~\ref{sec:synthesis}. 
To encode the data, the library was first amplified by 
limited-cycle PCR, and then sequenced on a single
lane of an Illumina HiSeq system, as described in Section~\ref{sec:sequencing}. Because synthesis and sequencing errors occurred with low frequency, the DNA blocks were correctly 
decoded using their own encodings and decoded copies of overlapping blocks. As a result, only $10$ bit errors were observed within 
the $5.27$ million encoded bits, i.e., the reported system error rate was less than $2 \times 10^{-6}$. 

\begin{figure*}
\begin{centering}
\includegraphics[scale=0.6]{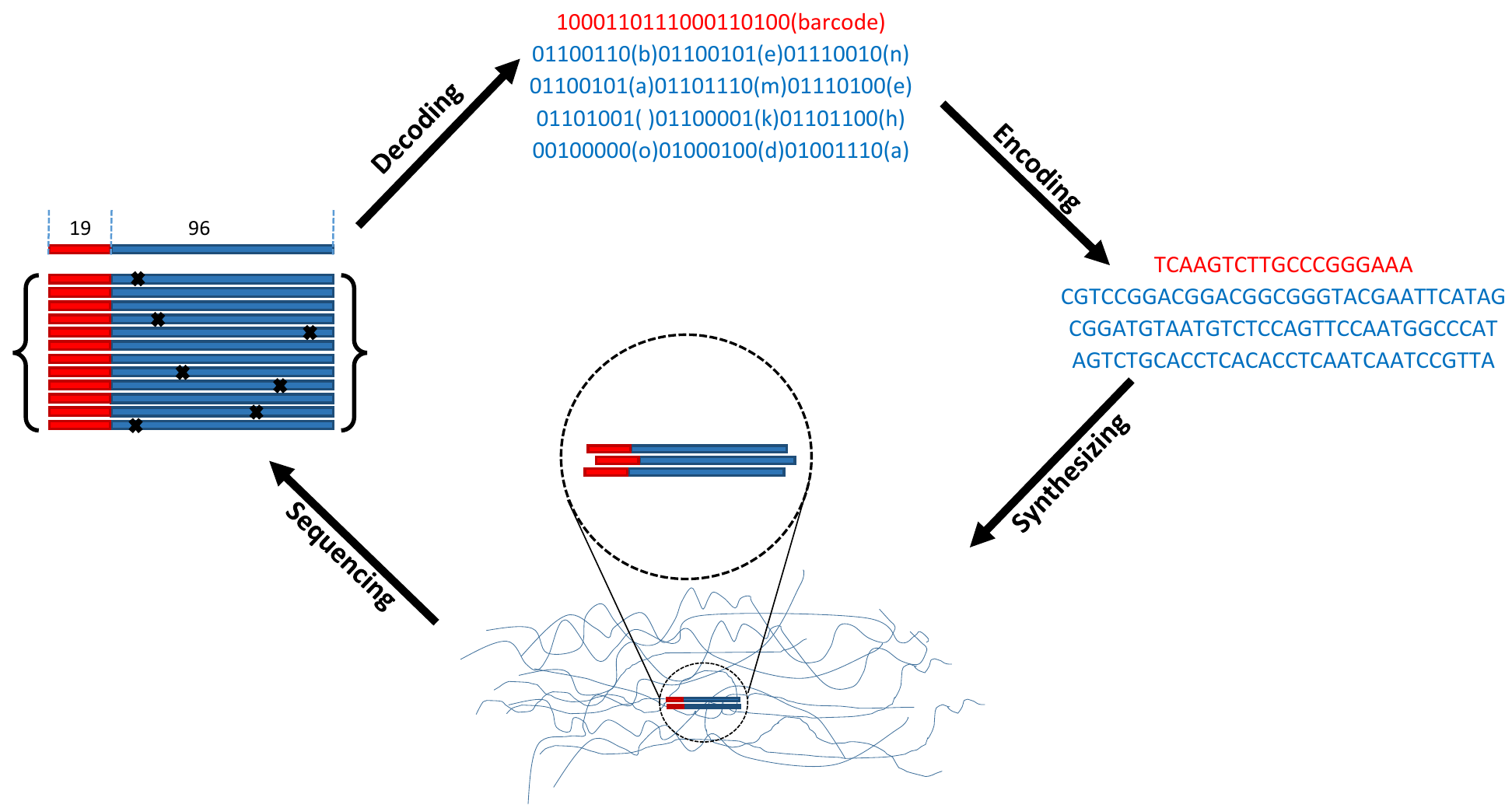} \label{fig:church}

\par\end{centering}
\protect\caption{A chosen text file is converted to ASCII format using $8$ bits, for each
symbol. Blocks of bits are subsequently 
encoded into DNA using a $1$ bit-per-oligonucleotide encoding. The entire $5.27$ Mb html file amounted to
$54,898$ oligonucleotides and was synthesized and eluted from a DNA
microchip. After amplification -- common primer sequences of the blocks are not shown -- 
the library was sequenced using an Illumina platform. Individual reads with the correct barcode and length were screened for consensus, and then converted back into bits comprising the
original file.}
\end{figure*}

The architecture of the Church-Gao-Kosuri DNA-based encoding system is illustrated in Figure~\ref{fig:church}.

\subsubsection*{Encoding example}
We provide next an example for the encoding algorithm proposed by Church-Gao-Kosuri~\cite{church2012next}. 
The text of choice is ``ferential DN''.
\begin{itemize}
\item First, each symbol is converted into its $8$ bit ASCII format. The encoding results in a 
binary string of length $12\times8=96$ of the following form: 
\begin{flalign*}
&\overset{\textrm{f}}{\overbrace{01100110}}\overset{\textrm{e}}{\overbrace{01100101}}
\overset{\textrm{r}}{\overbrace{01110010}}\overset{\textrm{e}}{\overbrace{01100101}}
\overset{\textrm{n}}{\overbrace{01101110}}\overset{\textrm{t}}{\overbrace{01110100}}&\\
&\overset{\textrm{i}}{\overbrace{01101001}}\overset{\textrm{a}}{\overbrace{01100001}}
\overset{\textrm{l}}{\overbrace{01101100}}\overset{\textrm{(space)}}{\overbrace{00100000}}
\overset{\textrm{D}}{\overbrace{01000100}}\overset{\textrm{N}}{\overbrace{01001110.}}&
\end{flalign*}

\item Second, a unique $19$ bits barcode is appended to the binary string
for the purpose of DNA block identification: here, we assume that the barcode
is $1000110111000110100$. This results in a binary string of length
$19+96=115,$ namely: 
\begin{flalign*}
&\overset{\textrm{barcode}}{\overbrace{1000110111000110100}}011001100110010101110010011&\\
&0010101101110011101000110100101100001011011000&\\
&01000000100010001001110.&
\end{flalign*}

\item Third, every bit $0$ is converted into $\mathtt{A}$ or $\mathtt{C}$
and every bit $1$ into $\mathtt{T}$ or $\mathtt{G}$. This conversion is performed randomly,
while disallowing homopolymer runs of length greater than three. The scheme also asks 
for balancing the GC content and controlling the secondary structure. For instance,
the following DNA code generated from the example binary text satisfies all
the aforementioned conditions:  
\begin{flalign*}
&\mathtt{TAACGTCTTGCCCGGAGAAATGAATTCATTCATATATGTCAGAA}&\\
&\mathtt{TTCATAGCGGATGTAATGTCTACGTCTCATAGGCCCATAGTCTG}&\\
&\mathtt{CCACTACACCATACATAACTCCGTTA.}&
\end{flalign*}

\item Finally, two primers of length $22$ nt are added to both
ends of the DNA block. The forward primer is $\mathtt{CTACACGACGCTCTTCCGATCT}$, while 
the backward primer is just the reverse complement of the forward
primer, $\mathtt{AGATCGGAAGAGCGGTTCAGCA}$. Hence, the encoded DNA codeword is
of length $22+115+22=159$ nt, and reads as:
\begin{flalign*}
&\overset{\textrm{forward}}{\overbrace{\mathtt{CTACACGACGCTCTTCCGATCT}}}\mathtt{TAACGTCTTGCCCGGAGAAATG}&\\
&\mathtt{AATTCATTCATATATGTCAGAATTCATAGCGGATGTAATGTCTA}&\\
&\mathtt{CGTCTCATAGGCCCATAGTCTGCCACTACACCATACATAACTCC}&\\
&\mathtt{GTTA}\underset{\textrm{backward}}{\underbrace{\mathtt{AGATCGGAAGAGCGGTTCAGCA.}}}&
\end{flalign*}
\end{itemize}

\subsection{The Goldman et al. Method}

To encode the digital information into a DNA sequence, Goldman \etal~\cite{goldman2013towards} 
started with a binary data set. The binary file representation
was obtained via ASCII encoding, using one byte per symbol (Step A). Each byte was subsequently 
converted into $5$ or $6$ trits via an optimal Huffman code for the underlying distribution
of the particular dataset used. The compressed file comprised $5.2\times10^{6}$ information bits (Step B). Each trit was then used to select one out of three DNA oligonucleotides 
differing from the last encoded oligonucleotide. This form of differential coding ensures that there are no homopolymer runs of any
length greater than one (Step C). Finally, the resulting DNA string was partitioned into segments of length $100$ oligonucleotides, 
each of which has the property that it overlaps in $75$ bases with each adjacent segment (Step D). 
This overlap ensures 4x coverage for each base. 
In addition, alternate segments of length $100$ were reverse complemented. Indexing information,
along with $2$ trits for file identification, $12$ trits for intra-file location
information (which can be used to encode up to $3^{14}$ unique segment locations), one parity-check
and one additional base are appended to both ends to indicate whether
the entire fragment was reverse complemented or not. The resulting fragment lengths of the constituent encodings amounted to $153,335$ oligos of length $117$.

As an experiment, Goldman \etal~\cite{goldman2013towards} encoded a digital data file of size $739$ KB with an estimated Shannon information of $5.2\times10^{6}$ bits into DNA. Their file included all $154$ of Shakespeare\textquoteright s
sonnets (ASCII text), a classic scientific paper (PDF format), a medium-resolution
color photograph of the European Bioinformatics Institute (JPEG $2000$
format), and a $26$-s excerpt from Martin Luther King\textquoteright s
$1963$ \textquoteleft I have a dream\textquoteright{} speech (MP3
format). The encoded strings were synthesized by an updated version of Agilent Technologies.
For each sequence, $1.2\times10^{7}$ copies were created, with $1$ base
error per $500$ bases, and sequenced on an Illumina HiSeq $2000$ system, and
decoded successfully. After several postprocessing steps, the original data was decoded with $100\%$ accuracy. 

\begin{figure*}

\begin{centering}
\includegraphics[scale=0.6]{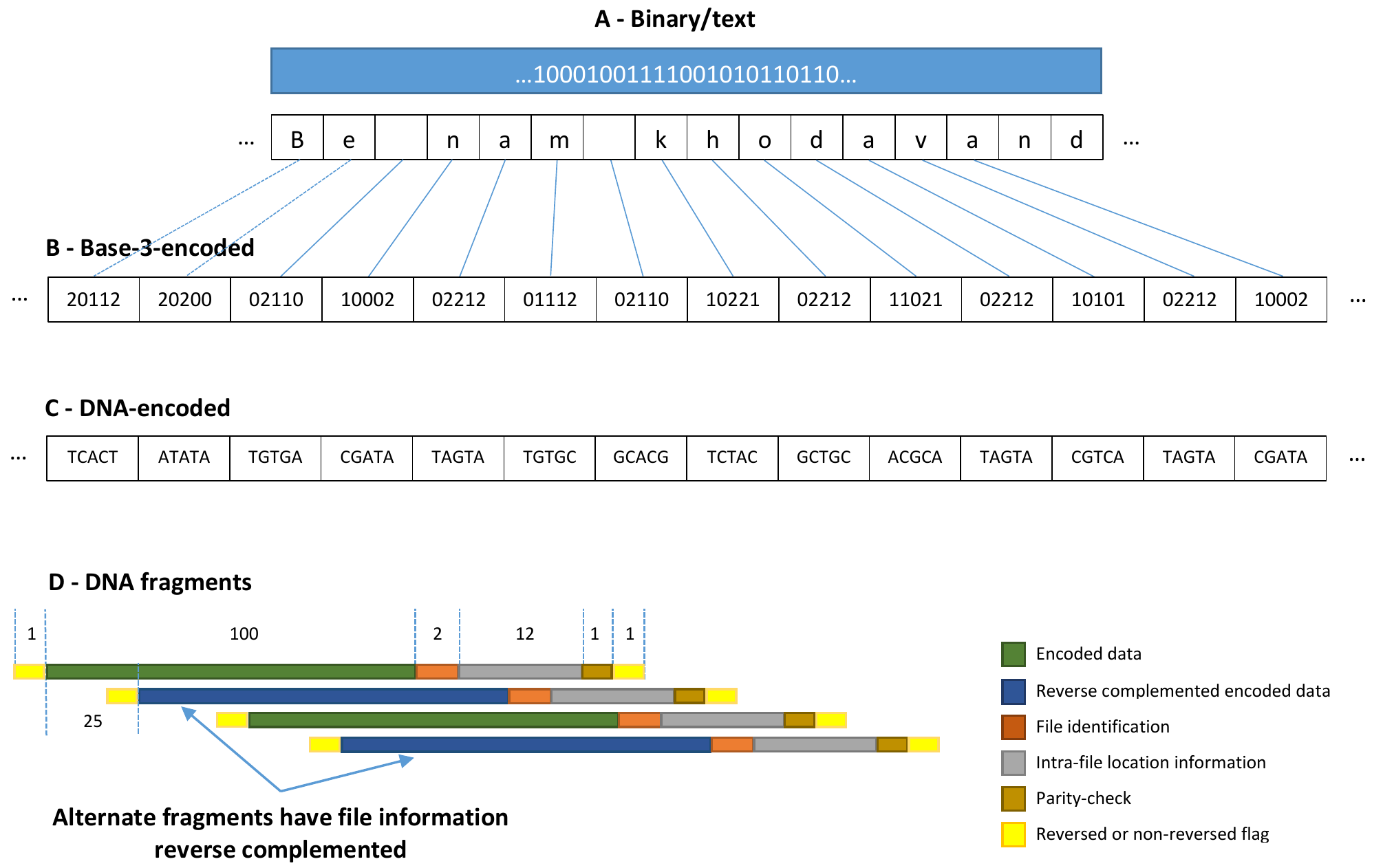} \label{fig:goldman}
\par\end{centering}

\protect\caption{The Goldman \etal encoding method using ASCII and differential coding, Huffman compression, four-fold coverage,
reverse complementation of alternate data blocks and single parity-check coding.}
\end{figure*}

The architecture of the Goldman \etal~DNA-based encoding system is illustrated in Figure~\ref{fig:goldman}.

\begin{figure*}
\begin{centering}
\includegraphics[scale=0.6]{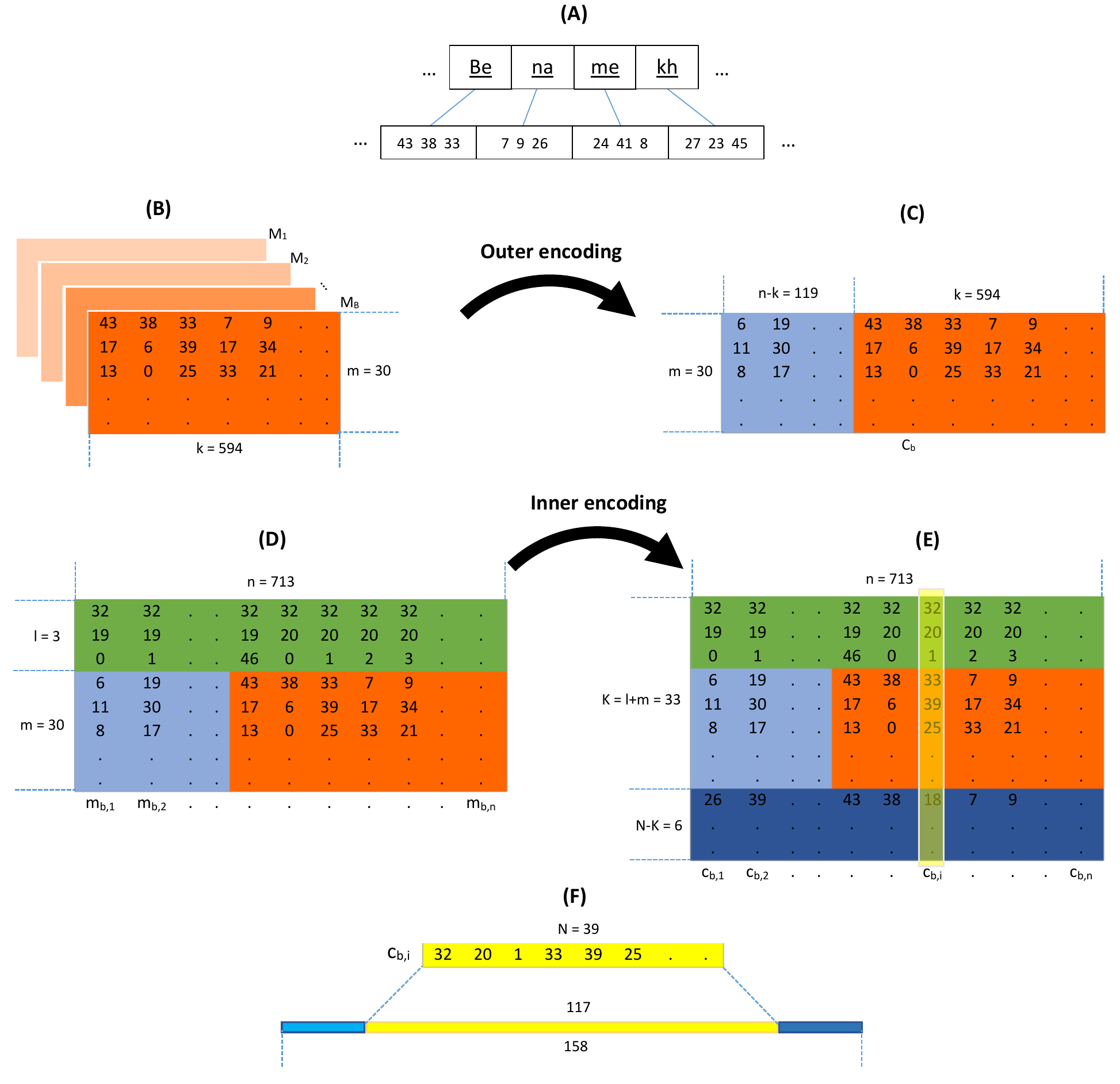} \label{fig:swiss}
\par\end{centering}
\protect\caption{The Grass \etal DNA text conversion, arraying (grouping) and encoding method.}
\end{figure*}

\subsubsection*{Encoding example}
We present next a short example of the encoding algorithm introduced by Goldman \etal~\cite{goldman2013towards}. The text to be encoded is ``Birney
and Goldman''.
\begin{itemize}
\item First, we apply Huffman coding base $3$ to compress the data, resulting in
\begin{flalign*}
&S_{1}=\overset{\textrm{B}}{\overbrace{20100}}\overset{\textrm{i}}{\overbrace{20210}}\overset{\textrm{r}}{\overbrace{10101}}\overset{\textrm{n}}{\overbrace{00021}}\overset{\textrm{e}}{\overbrace{20001}}\overset{\textrm{y}}{\overbrace{222111}}\overset{\textrm{(space)}}
{\overbrace{02212}}\overset{\textrm{a}}{\overbrace{01112}}&\\
&\overset{\textrm{n}}{\overbrace{00021}}\overset{\textrm{d}}{\overbrace{22100}}\overset{\textrm{(space)}}{\overbrace{02212}}\overset{\textrm{G}}{\overbrace{222212}}\overset{\textrm{o}}
{\overbrace{02110}}\overset{\textrm{l}}{\overbrace{02101}}\overset{\textrm{d}}{\overbrace{22100}}\overset{\textrm{m}}{\overbrace{11021}}\overset{\textrm{a}}{\overbrace{01112}}&\\
&\overset{\textrm{n}}{\overbrace{00021.}}&
\end{flalign*}

\item Let $n=len\left(S_{1}\right)=92,$ which equals $10102$ in base $3$.
Hence, we set $S_{2}=00000000000000010102$ (an encoding of length $20$) and $S_{3}=0000000000000$
(an encoding of length $13$). Therefore, 
\begin{flalign*}
&S_{4}=S_{1}S_{3}S_{2}=201002021010101000212000122211102&\\
&21201112000212210002212222212021100210122100110&\\
&210111200021000000000000000000000000000010102,&
\end{flalign*}
of total length $92+13+20=125$.
\item Applying differential coding to $S_{5}$ according to the table
\[
\begin{array}{cccc}
 &  & \textrm{next}\\
\textrm{previous} & 0 & 1 & 2\\
\mathtt{A} & \mathtt{C} & \mathtt{G} & \mathtt{T}\\
\mathtt{C} & \mathtt{G} & \mathtt{T} & \mathtt{A}\\
\mathtt{G} & \mathtt{T} & \mathtt{A} & \mathtt{C}\\
\mathtt{T} & \mathtt{A} & \mathtt{C} & \mathtt{G}
\end{array}
\]
results in an encoding of $S_{4}$ that reads as
\begin{flalign*}
&S_{5}=\mathtt{TAGTATATCGACTAGTACAGCGTAGCATCTCGCAGCGAGAT}&\\
&\mathtt{ACGCTGCTACGCAGCATGCTGTGAGTATCGATGACGAGTGACTCT}&\\
&\mathtt{GTACAGTACGTACGTACGTACGTACGTACGTACGACTAT.}&
\end{flalign*}

\item Since $len\left(S_{5}\right)=125$, there are two DNA blocks
$F_{0}$ and $F_{1}$ of length $100$ overlapping in exactly $75$ bps, i.e.,
\begin{flalign*}
&F_{0}=\mathtt{TAGTATATCGACTAGTACAGCGTAGCATCTCGCAGCGAGAT}&\\
&\mathtt{ACGCTGCTACGCAGCATGCTGTGAGTATCGATGACGAGTGACTCT}&\\
&\mathtt{GTACAGTACGTACG}&
\end{flalign*}
 and
\begin{flalign*}
&F_{1}=\mathtt{CATCTCGCAGCGAGATACGCTGCTACGCAGCATGCTGTGAG}&\\
&\mathtt{TATCGATGACGAGTGACTCTGTACAGTACGTACGTACGTACGTAC}&\\
&\mathtt{GTACGTACGACTAT.}&
\end{flalign*} 
Moreover, the odd-numbered DNA blocks are reverse complemented so that
\begin{flalign*}
&F_{1}=\mathtt{ATAGTCGTACGTACGTACGTACGTACGTACGTACTGTACAG}&\\
&\mathtt{AGTCACTCGTCATCGATACTCACAGCATGCTGCGTAGCAGCGTAT}&\\
&\mathtt{CTCGCTGCGAGATG.}&
\end{flalign*} 

\item The file identification for the text equals $12$. This gives $ID_{0}=ID_{1}=12$.
The $12$ trits intra-file location for $F_{0}$ equals $intra_{0}=000000000000$
and for $F_{1}$, it equals $intra_{1}=000000000001$. The parity check $P_{i}$
for block $F_{i}$ is the sum of the bits at odd locations in $ID_{i}intra_{i}$ taken mod $3$. Thus, $P_{0}=P_{1}=1+0+0+0+0+0+0\overset{\textrm{mod }3}{\equiv}1$. By appending $IX_{i}=ID_{i}intra_{i}P_{i}$ to $F_{i}$ we get,
\begin{flalign*}
&F_{\ensuremath{0}}^{\prime}=\mathtt{TAGTATATCGACTAGTACAGCGTAGCATCTCGCAGCGAGA}&\\
&\mathtt{TACGCTGCTACGCAGCATGCTGTGAGTATCGATGACGAGTGACT}&\\
&\mathtt{CTGTACAGTACGTACG\:AT\:ACGTACGTACGT\:C}&
\end{flalign*} 
 and 
\begin{flalign*}
&F_{\ensuremath{1}}^{\prime}=\mathtt{ATAGTCGTACGTACGTACGTACGTACGTACGTACTGTACA}&\\
&\mathtt{GAGTCACTCGTCATCGATACTCACAGCATGCTGCGTAGCAGCGT}&\\
&\mathtt{ATCTCGCTGCGAGATG\:AT\:ACGTACGTACGA\:G.}&
\end{flalign*} 
\item In the last step, we prepend $\mathtt{A}$ or $\mathtt{T}$ and append $\mathtt{C}$
or $\mathtt{G}$ to even and odd blocks, respectively. The resulting DNA codewords equals
\begin{flalign*}
&F_{\ensuremath{0}}^{\prime\prime}=\mathtt{A\:TAGTATATCGACTAGTACAGCGTAGCATCTCGCAGCGA}&\\
&\mathtt{GATACGCTGCTACGCAGCATGCTGTGAGTATCGATGACGAGTGA}&\\
&\mathtt{CTCTGTACAGTACGTACGATACGTACGTACGTC\:G}&
\end{flalign*} 
 and 
\begin{flalign*}
&F_{\ensuremath{1}}^{\prime\prime}=\mathtt{T\:ATAGTCGTACGTACGTACGTACGTACGTACGTACTGTA}&\\
&\mathtt{CAGAGTCACTCGTCATCGATACTCACAGCATGCTGCGTAGCAGC}&\\
&\mathtt{GTATCTCGCTGCGAGATGATACGTACGTACGAG\:C.}&
\end{flalign*}
\end{itemize}

\subsection{The Grass et al. Method}

As it is apparent from the previous exposition, the Church-Gao-Kosuri and Goldman \etal methods did 
not implement error-correction schemes that go beyond single parity-check coding of fragments. Additional error-correction was accomplished via four-fold coverage. Nevertheless, with the relatively low synthesis and sequencing accuracies of the proposed platforms, the lack of advanced error-correction solutions may be a significant disadvantage. Furthermore, additional errors may arise due to ``aging'' of the media, as there are no best practices for physically storing the DNA strings to maximize their stability over long periods of time.

In~\cite{grass2015robust}, the authors addressed both these issues by implementing a specialized error-correcting scheme and by outlining best practices for DNA media maintainance. Their experiments
show that by only combining these two approaches, one should be able to store and recover information encoded in the DNA from the Global Seed Vault (at $18$ $8$C) for hundreds of thousands of years.

The steps applied in~\cite{grass2015robust} for encoding text onto DNA include:
\begin{itemize}
\item \emph{Grouping}: Every two characters are mapped to tree elements in $\mathbb{F}\left(47\right)$, the finite field of size $47$, via base conversion from $256^{2}$ to $47^{3}$. This results in $B$ information arrays of dimension $m\times k$ information blocks, with elements in $\mathbb{F}\left(47\right)$. The information arrays are denoted by $M_{b},$ with $b\in\left\{ 1,\ldots,B\right\}$ (see Figure~\ref{fig:swiss} for the notation and for an illustration of the grouping). Hence,
each block $M_{b}$ corresponds to a vector of length $k$, with elements
in $\mathbb{F}\left(47^{m}\right)$.
\item \emph{Outer Encoding}: Each block $M_{b}$ is encoded using a Reed-Solomon (RS) code
over $\mathbb{F}\left(47^{m}\right)$ to a codeword $C_{b}$ of length $n$. 
This encoding procedure leads to blocks of size $m\times n$. To uniquely identify
each column in $C_{b}$, one has to append $l$ elements in $\mathbb{F}\left(47\right)$
to each column. This produces vectors $m_{b,1},\ldots,m_{b,n}$ of
length $K=l+m$ each.
\item \emph{Inner Encoding}: Each vector $m_{b,i}$ is mapped to a vector of length
$N$ over $\mathbb{F}\left(47\right)$ by using RS coding to obtain the codewords $c_{b,i}$.
\item \emph{Mapping to DNA Strings}: Each element in $c_{b,i}$ is converted to
a DNA string of length $3$ so that no homopolymers of length three or longer appear. This process results in a DNA strings of length $3N$. To complete the mapping and encoding, two fixed primers are attached to both ends of each created DNA string and used for rapid sequencing. 
\end{itemize}

To experimentally test their method, the authors started with $83$KB of uncompressed text containing the Swiss Federal Charter from $1291$ and the English
translation of the Methods of Archimedes. This information was encoded into $4991$ DNA oligos of length $158$. Each of the oligostrings comprised $117$ ``information'' nucleotides. The sequences were synthesized using the CustomArray electrochemical microarray technology described in the previous sections, with a total price of $2,500$ USD. In the process of information retrieval, custom PCR was combined with sequencing on the Illumina MiSeq platform.

The individual decay rates of different DNA strands are mostly influenced by the storage temperature and the water concentration of the DNA storage environment. Four different dry storage technologies for DNA were tested: pure solid-state DNA, DNA on a Whatman FTA filter card, DNA on a biopolymeric storage matrix and DNA encapsulated in silica. Among the tested methods, DNA encapsulated in silica appears to offer the most durable storage format, as silica
has the lowest water concentration and it separates DNA molecules
from the environment through an inorganic layer. Therefore, the quality of preservation
is not affected by environmental humidity, which is important since unlike low temperature (e.g. permafrost) and absence of light, humidity is relatively hard to control. DNA storage systems within silica substrates have the further advantages of exceptional stability against oxidation and photoresistance, provided that an additional titania layer is added to silica.

\section{Random Access and Rewritable DNA-Based Storage} \label{sec:coding}

Although the techniques described in~\cite{church2012next,goldman2013towards,grass2015robust} 
provided a number of solutions for DNA storage, they did not address one important issue: accurate partial
and random access to data. In all the cited methods, one has to reconstruct
the whole text in order to read or retrieve the information encoded
even in a few bases, as the addressing methods used only allow
for determining the position of a read in a file, but cannot ensure
precise selection of reads of interest due to potential undesired cross-hybridization
between the primers and parts of the information blocks. 
Moreover, all current designs support read-only storage. 
Adapting the archival storage solutions to address random access and rewriting appears complicated, due to the storage format that involves reads of length $100$ bps shifted by $25$ bps so as to
ensure four-fold coverage of the sequence. In order
to rewrite one base, one needs to selectively access and modify four consecutive reads.

The drawbacks of the archival architectures were addressed in~\cite{yazdi2015rewritable}, where new coding-theoretic
methods were introduced to allow for rewriting and controlled random access.
 \vspace{-0.1in}
\subsection{The Yazdi et al. Method}

To overcome the aforementioned issues, Yazdi \etal~\cite{yazdi2015rewritable} developed
a new, random-access and rewritable DNA-based storage architecture
based on DNA sequences endowed with specialized address strings that
may be used for selective information access and encoding with inherent
error-correction capabilities. The addresses are designed to be mutually
uncorrelated, which means that for a set of addresses $\A=\left\{ \va_{1},\ldots,\va_{M}\right\}$, 
each of length $n$, and any two distinct addresses $\va_{i},\va_{j}\in \A$, no prefix
of $\va_{i}$ of length $\leq n-1$ appears as a proper suffix of $\va_{j}$. 

\begin{figure}
\begin{centering}
\includegraphics[scale=1.35]{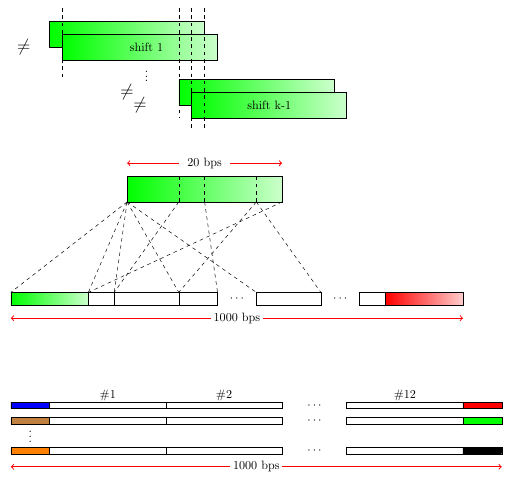} \label{fig:rewrite}

\par\end{centering}
\protect\caption{Data format and encoding for the random access, rewritable architecture of~\cite{yazdi2015rewritable}.}
\end{figure}

Information is encoded into DNA blocks of length $L=2n+ml$. The $i$th
block, $B_{i}$, is flanked at both ends by two unique addresses, one of which, say 
$\va_{i}$, of length $n$, is used for encoding. The remainder of the block is divided into
$m$ sub-blocks $sub_{i,1},\ldots,sub_{i,m},$ each of length $l$. Encoding of the 
block $B_{i}$ is performed by first dividing the classical digital information stream into $m$ 
non-overlapping segments and then mapping them to integers $x_{1},\ldots,x_{m}$, respectively. 
Then, each $x_{j}$, for $1\leq j\leq m$, is encoded into a DNA sub-block $sub_{i,j}$ of length $l$ using an algorithm, named $\textsc{Encode}_{\va_{i},l}(x_{j}),$ introduced in~\cite{yazdi2015rewritable} and described in detail in the next section. The algorithm represents an extension of \emph{prefix-synchronized
coding} methods~\cite{gilbert1960synchronization} (see Figure~\ref{fig:rewrite} for an illustration). 
Given that the addresses in $A$ are chosen to be mutually uncorrelated and at large Hamming distance from each other, no $a_{i}$ appears as a subword in any DNA block, except at one flanking end of the $i$th block.
This feature enables highly sensitive random access and accurate rewriting using the DNA editing techniques described in Section~\ref{sec:editing}.

To experimentally test their scheme, Yazdi \etal~\cite{yazdi2015rewritable} used the introductory pages 
of five universities retrieved from Wikipedia, amounting to a total size of $17$KB in ASCII format. The 
text was encoded into $32$ DNA blocks of length
$L=1000$ bps. To facilitate addressing, they constructed a set of $32$ pairs of mutually uncorrelated addresses and used $32$ of them for encoding. The addresses used for encoding $\A=\left\{ \va_{1},\ldots,\va_{32}\right\}$ were each of length $n=20$ bps. Different words in the text were counted and tabulated in a dictionary. Each word in the dictionary was converted into a binary sequence of length $24$. Groups of six consecutive words in the file were grouped and mapped to binary strings of length $6\times24=144$. Two bits $11$ were appended to the left
hand side of each binary sequence of length $144$ to shift the range of encoded values, resulting in sequences of length $146$ bits. The binary sequences were then translated into DNA
sub-blocks of length $l=80$ bps using $\textsc{Encode}_{(\cdot)}(\cdot)$.
Next, $m=12$ sub-blocks of length $80$ bps each were adjoined to form
a DNA string of length $12\times80=960$ bps. To complete the encoding, each string
of length $960$ bps was equipped with two unique primers of length $20$
bps at its ends, forming a DNA block of length $L=20+960+20=1000$
bps\footnote{Two different addresses were used to terminate one sequence because of DNA synthesis issues, as having one long repeated string at both flaking ends lead to undesired secondary structures.}. The resulting DNA sequences were synthesized by IDT~\cite{idtdna}, at the price of \$149 per $1000$ bps.

To test the rewriting method, all $32$ linear $1000$ bps fragments
were mixed, and the information in three blocks was rewritten in the
DNA encoded domain using both gBlocks and OEPCR editing techniques, described in Section~\ref{sec:editing}. The rewritten blocks were selected, amplified and Sanger sequenced to verify that selection and rewriting were performed with 100\% accuracy.

\subsubsection*{Encoding example}
We illustrate next the encoding and decoding procedure described in~\cite{yazdi2015rewritable} for the short address string 
$\va=\mathtt{ACCTG}$, which can easily be verified to be self-uncorrelated (i.e., no prefix of the sequence equals a suffix of the sequence). For the sequence of integers $G_{n,1},G_{n,2},\ldots,G_{n,7}$, the construction of which will be described in detail in~\ref{sec:prefix}, one can verify that
\[
\left(G_{n,1},G_{n,2},\ldots,G_{n,7}\right)=\left(3,9,27,81,267,849,2715\right).
\]
Here, $n$ denotes the length of the address string, which in this case equals five. 
The algorithm $\textsc{Encode}_{\va,8}(550)$ produces
\begin{flalign*}
    &550=0\times G_{5,7}+550& \\
    &\Rightarrow\textsc{Encode}_{\va,8}(550)=\underline{\mathtt{C}}\textsc{Encode}_{\va,7}(550) & \\
    &550=0\times G_{5,6}+550&\\
    &\Rightarrow\textsc{Encode}_{\va,7}(550)=\underline{\mathtt{C}}\textsc{Encode}_{\va,6}(550) &\\ 
    &550=2\times G_{5,5}+0\times G_{5,4}+16&\\
    &\Rightarrow\textsc{Encode}_{\va,6}(550)=\underline{\mathtt{AA}}\textsc{Encode}_{\va,4}(16), &\\
    &16=0\times3^{3}+1\times3^{2}+2\times3^{1}+1\times3^{0}&\\
    &\Rightarrow\textsc{Encode}_{\va,4}(16)=\underline{\mathtt{ATCT}}, &\\
    &\Rightarrow\textsc{Encode}_{\va,8}(550)=\underline{\mathtt{CCAAATCT}} &\\
\end{flalign*}
When running $\textsc{Decode}_{\va}(X)$ on the encoded output $X=\underline{\mathtt{CCAAATCT}}$, the following steps are
executed:

\begin{flalign*}
    &\Rightarrow\textsc{Decode}_{\va}(\mathtt{\underline{C}CAAATCT})=0\times G_{5,7}&\\
    &+\textsc{Decode}_{\va}(\mathtt{CAAATCT})&\\
    &\Rightarrow\textsc{Decode}_{\va}(\mathtt{\underline{C}AAATCT})=0\times G_{5,6} & \\
    &+\textsc{Decode}_{\va}(\mathtt{AAATCT}), & \\
    &\Rightarrow\textsc{Decode}_{\va}(\mathtt{\underline{AA}ATCT})=2\times G_{5,5}+0\times G_{5,4}&\\
    &+\textsc{Decode}_{\va}(\mathtt{ATCT})&\\
    &\Rightarrow\textsc{Decode}_{\va}(\mathtt{\underline{ATCT}})=16&\\ 
    &\Rightarrow\textsc{Decode}_{\va}(\mathtt{CCAAATCT})=2\times G_{5,5}+16=550&
\end{flalign*}

\subsection{Address Design and Constrained Coding} \label{sec:constrained}

To encode information on a DNA media, Yazdi \etal~\cite{yazdi2015rewritable} first designed a set $\A$ of address sequences, each of length $n$, 
that satisfies a number of constraints. These constraints make the codewords suitable for selective random access; given the address set $\A$, they also constructed a code $\C_\A(\ell)$ of length $\ell$ and 
provided efficient methods to encode and decode messages to codewords in $\C_\A(\ell)$.
In their experiment, Yazdi \etal{} chose $n=20$ and $\ell=80$ and stored
twelve data subblocks of length $80$, each corresponding to the codewords in $\C_\A(\ell)$, and flanked these subblocks with two address sequences to obtain a datablock of length $1000$ bps.

In Section~\ref{sec:constrained}, we describe the design constraints for the address sequences and 
relate these constraints to previously studied concepts such as {\em running digital sums}
and {\em sequence correlation}. In Section \ref{sec:prefix}, we describe the desired properties of $\C_\A(\ell)$ and present the encoding schemes developed by Yazdi \etal{} based on {\em prefix-synchronized schemes} described by Morita \etal \cite{morita1996construction}.

\subsection{Constrained Coding for Address Sequences} \label{sec:constrained}

Constrained coding serves two purposes in the design of address sequences.
First, it ensures that DNA patterns prone to sequencing errors are avoided. 
Second, it allows DNA blocks to be accurately accessed, amplified and selected without perturbing other blocks in the DNA pool. We remark that while these constraints apply to address primer design, 
they indirectly govern the properties of the fully encoded DNA information blocks.
%
%
Specifically, we require the address sequences to satisfy the following constraints:
\begin{enumerate}[(C1)]
\item {\em Constant GC content (close to $50\%$) for all the prefixes of the sequences
of sufficiently long length}. 
DNA strands with
$50\%$ GC content are more stable than DNA strands with lower or
higher GC content and have better coverage during sequencing. Since encoding user information is accomplished
via prefix-synchronization, it is important to impose the GC content constraint on the addresses as well as their prefixes, as the latter requirement ensures that all fragments of encoded data blocks are balanced as well.
Given $D> 0$, we define a sequence to be {\em $D$-GC-prefix-balanced} ($D$-GCPB) if 
for all prefixes (including the sequence itself), 
the difference between the number of $\dG$ and $\dC$ bases and the number of $\dA$ and $\dT$ bases
is at most $D$. A set of address sequences is $D$-GCPB if all sequences in the set are $D$-GCPB.

\item  {\em Large mutual Hamming distance}. This reduces the probability
of erroneous address selection. Recall that the Hamming distance between
two strings of equal length equals the number of positions at which the
corresponding symbols disagree.
Given $d>0$, we design our set of sequences such that the Hamming distance between any pair of distinct sequences
is at least d.
\item {\em Uncorrelatedness of the addresses}. This imposes the restriction that prefixes of one address do not appear as suffixes of the same or another address. The motivation for this new constraint comes from the fact that addresses are used to provide unique identities for the blocks, and that their substrings should therefore not appear in ``similar form'' within other addresses. Here, ``similarity'' is assessed in terms of hybridization affinity. Furthermore, long undesired prefix-suffix matches may lead to assembly errors in blocks during joint sequencing. Most importantly, uncorrelated sequences may be jointly avoided via simple and efficient coding methods. Hence, one can ensure that address sequences only appear at the flanking ends of the blocks and nowhere else in the encoding.
\item  {\em Absence of secondary (folding) structure for the address primers}. Such structures may cause errors in the process of PCR amplification and fragment rewriting. 
\end{enumerate}


As observed by Yazdi \etal{}, constructing addresses that simultaneously satisfy the constraints C1-C4 and 
determining bounds on the largest number of such sequences is prohibitively complex \cite{yazdi2015rewritable}.
To mitigate this problem, Yazdi \etal{} used a \emph{semi-constructive} address design approach, in which balanced error-correcting codes are designed independently, and subsequently expurgated so as to identify a large set of mutually uncorrelated sequences. The resulting sequences are subsequently tested for secondary structure using~\emph{mfold} and~\emph{Vienna}~\cite{rouillard2003oligoarray}. 

In the same paper, Yazdi \etal{} observed that if one considers the constraints individually or one focuses on certain proper subsets of constraints, it is possible to construct families of codes whose size grow exponentially with code length.
To demonstrate this, Yazdi \etal{} borrowed concepts from other areas in coding theory. We provide an overview of these techniques in what follows. 

\vspace{2mm}

\noindent{\em Running Digital Sums.} An important criteria for selecting block addresses is to ensure
that the corresponding DNA primer sequences have prefixes with a GC content approximately
equal to $50\%$, and that the sequences are at large pairwise Hamming distance. 
Due to their applications in optical storage, codes that address related issues have been studied in 
a slightly different form under the name of \emph{bounded running digital sum} (BRDS) codes~\cite{cohen1991dc,blaum1993error}. A detailed overview of this coding technique may be found in~\cite{cohen1991dc}.

Fix an integer $D>0$. A binary sequence $\va$ has a $D$-bounded running digital sum ($D$-BRDS) if
for any prefix of $\va$ (including $\va$ itself), the number of zeroes and the number of ones differ by at most $D$.
A set $\vA$ of binary sequences is $D$-BRDS if all sequences in $\vA$ have $D$-BRDS.
A $1$-BRDS set $\vA$ with minimum distance $2d$ may be obtained from a binary code with distance $d$ 
via the following theorem.  

\begin{thm}[{\cite[Thm 2]{blaum1993error}}] \label{rds-blaum}
If a binary unrestricted code of length $n$, size $M$ and minimum distance $d$ exists, 
then a $1$-BRDS set of length $2n$ and minimum distance $2d$ and size $M$ exists.
\end{thm}
Hence, it follows from the Gilbert-Varshamov bound that there exists a $1$-BRDS set 
of length $2n$ and minimum distance $2d$ whose size is at least $2^n/\left(\sum_{j=0}^{d-1} \binom{n}{j}\right)$.

A set of DNA sequences over $\left\{ \mathtt{A,T,G,C}\right\}$ may then be constructed in a straightforward manner by mapping each $0$ into one of the bases $\left\{ \mathtt{A,T}\right\},$ and $1$ into one of the bases $\left\{ \mathtt{G,C}\right\}$. 
In other words, a $D$-BRDS set of length $n$ and size $M$ yields a $D$-GCPB set of sequences of size $M$.
For $0<d\le n$, $D>0$, let $M_1(n,d;D)$ denote the maximum size of a $D$-GCPB set of sequences of length $n$ and minimum distance $d$. Furthermore, for $q>0$, let $A_q(n,d)$ denote the maximum size of a $q$-ary code with minimum distance $d$.

Applying Theorem~\ref{rds-blaum} and the simple mapping above, we have the following estimates for the size of codes satisfying C1 and C2.
\begin{thm}\label{thm:M1}
Fix $0<d\le n$, $D=1$. Then
\begin{equation}
A_2(n/2,d/2) \le M_1(n,d;1) \le A_4(n,d).
\end{equation}
\end{thm}

\vspace{4mm}

\noindent {\em Sequence Correlation}

We describe next the notion of autocorrelation of a sequence 
and introduce the related notion of mutual correlation of sequences. 
It was shown in \cite{guibas1978maximal} that the
autocorrelation function is the crucial mathematical concept for studying
sequences avoiding forbidden words (strings) and subwords (substrings). In order to accommodate
the need for selective retrieval of a DNA block without accidentally selecting any undesirable blocks, we find it necessary to also introduce the notion of mutually uncorrelated sequences. 

Let $X$ and $Y$ be two words, possibly of different lengths, over
some alphabet of size $q>1$. The correlation of $X$ and $Y$, denoted
by $X\circ Y$, is a binary string of the same length as $X$. The
$i$-th bit (from the left) of $X\circ Y$ is determined by placing
$Y$ under $X$ so that the leftmost character of $Y$ is under the
$i$-th character (from the left) of $X$, and checking whether the characters
in the overlapping segments of $X$ and $Y$ are identical. If they
are identical, the $i$-th bit of $X\circ Y$ is set to $1$, otherwise,
it is set to $0$. For example, for $X=\mathtt{GTAGTAG}$ and $Y=\mathtt{TAGTAGCC}$, $X\circ Y=0100100$, as depicted below.

Note that in general, $X\circ Y\neq Y\circ X$, and that the two correlation
vectors may be of different lengths. In the example above, we have
$Y\circ X=00000000$. The autocorrelation of a word $X$ equals $X\circ X$.

In the example below, $X\circ X=1001001$.

\[
\begin{array}{cccccccccccccccc}
X= & \mathtt{G} & \mathtt{T} & \mathtt{A} & \mathtt{G} & \mathtt{T} & \mathtt{A} & \mathtt{G}\\
Y= & \mathtt{T} & \mathtt{A} & \mathtt{G} & \mathtt{T} & \mathtt{A} & \mathtt{G} & \mathtt{C} & \mathtt{C} &  &  &  &  &  &  & 0\\
 &  & \mathtt{T} & \mathtt{A} & \mathtt{G} & \mathtt{T} & \mathtt{A} & \mathtt{G} & \mathtt{C} & \mathtt{C} &  &  &  &  &  & 1\\
 &  &  & \mathtt{T} & \mathtt{A} & \mathtt{G} & \mathtt{T} & \mathtt{A} & \mathtt{G} & \mathtt{C} & \mathtt{C} &  &  &  &  & 0\\
 &  &  &  & \mathtt{T} & \mathtt{A} & \mathtt{G} & \mathtt{T} & \mathtt{A} & \mathtt{G} & \mathtt{C} & \mathtt{C} &  &  &  & 0\\
 &  &  &  &  & \mathtt{T} & \mathtt{A} & \mathtt{G} & \mathtt{T} & \mathtt{A} & \mathtt{G} & \mathtt{C} & \mathtt{C} &  &  & 1\\
 &  &  &  &  &  & \mathtt{T} & \mathtt{A} & \mathtt{G} & \mathtt{T} & \mathtt{A} & \mathtt{G} & \mathtt{C} & \mathtt{C} &  & 0\\
 &  &  &  &  &  &  & \mathtt{T} & \mathtt{A} & \mathtt{G} & \mathtt{T} & \mathtt{A} & \mathtt{G} & \mathtt{C} & \mathtt{C} & 0
\end{array}
\]

\begin{defn}
\emph{A sequence $X$ is self-uncorrelated if} $X\circ X=10\ldots0$. A set of sequences $\{{X_1,X_2,\ldots,X_m\}}$ is
termed mutually uncorrelated if each sequence is self-uncorrelated and if all pairs of distinct sequences satisfy 
$X_i\circ X_j=0\ldots0$ and $X_j\circ X_i=0\ldots0$.
\end{defn}
The notion of mutual uncorrelatedness may be relaxed by requiring that 
only sufficiently long prefixes do not match sufficiently long suffixes of other sequences. Sequences with this property, and at sufficiently large Hamming distance, eliminate undesired address cross-hybridization during selection.

Mutually uncorrelated codes were studied by many authors under a variety of names.
Levenshtein first introduced them in 1964 under the name `strongly regular codes' \cite{levenshtein1964decoding}, 
suggesting that the codes are interesting for synchronisation applications.
Inspired by the use of distributed sequences in frame synchronisation applications by van Wijngaarden and Willink \cite{de2000frame}, 
Baji\'c and Stojanovi\'c \cite{bajic2004distributed} recently independently rediscovered mutually uncorrelated codes using the term 'cross-bifix-free'  
(see also~\cite{bilotta2012new, chee2013cross, blackburn2013non} for recent papers and the references therein). The maximum size of a set of mutually uncorrelated code has been determined up to a constant factor by Blackburn~\cite{blackburn2013non}. We state his result below.

\begin{thm}
Let $M_2(n)$ be the maximum size of a set of mutually uncorrelated sequences of length $n$. 
Then
\label{thm:M2}
\[
\frac{3\cdot 4^n}{4en}(1-o(1)) \leq M_2(n) \le \frac{4^n}{n}\left(1-\frac 1n\right)^{n-1}=\frac{4^n}{en}(1+o(1)).
\]
\end{thm}

We point to an interesting construction by Bilotta \etal{} \cite{bilotta2012new} and 
provide a simple modification to obtain a set of sequences satisfying C1 and C3.
To do so, we introduce a simple combinatorial object called a {\em Dyck word}.
A Dyck word is a binary string consisting of $m$ zeroes and $m$ ones such that no prefix of the word has more zeroes than ones.

By definition, a Dyck word necessarily starts with a one and ends with a zero.
Consider a set $\D$ of Dyck words of length $2m$ and define the following set of words of length $2m+1$,
\[\A\triangleq\{1\va: \va\in\D\}.\] 
Bilotta \etal{} demonstrated that $\A$ is a mutually uncorrelated set of sequences.

A Dyck word has height at most $D$ if for any prefix of the word, the difference between the number of ones and 
the number of zeroes is at most $D$. In other words, a Dyck word has height at most $D$ if it has $D$-BRDS.
Let Dyck$(m,D)$ denote the number of Dyck words of length $2m$ and height at most $D$. de Bruijn \etal{} \cite{bruijn1972average} proved that 
for fixed values of $D$, 
\begin{equation}
{\rm Dyck}(m,D)\sim\frac{4^m}{D+1}\tan^2 \frac{\pi}{D+1}\cos^{2m}\frac{\pi}{D+1}.
\end{equation}
Here, $f(m)\sim g(m)$ means that $\lim_{m\to\infty} f(m)/g(m)=1$.

As with Bilotta \etal{}, we observe that if we prepend Dyck words of length $2m$ and height at most $D$ by $1$, we obtain a mutually uncorrelated $D+1$-BRDS set of binary words of length $2m+1$. 
As before, we map 0 and 1 into $\{\dA,\dT\}$ and $\{\dC,\dG\}$, respectively, and 
obtain a mutually uncorrelated $D+1$-GCPB set of sequences.

\begin{thm}
Let $M_3(n,D)$ be the maximum size of a mutually uncorrelated $D$-GCPB set of sequences of length $n$. 
If $n$ is odd and $D\ge 2$, then
\label{thm:M3}
\[
M_3(n,D) \ge \frac{2^{n-1}}{D}\tan^2 \frac{\pi}{D}\cos^{n-1}\frac{\pi}{D}(1+o(1)).
\]
\end{thm}
 
As already pointed out, it is an open problem to determine the largest number of address sequences that jointly satisfy the constraints C1 to C4. 
We conjecture that the number of such sequences is exponential in $n$,
 since the number of words that satisfy C1+C2, C3, and C1+C3 separately is exponential (see Theorems \ref{thm:M1},\ref{thm:M2}, \ref{thm:M3}). Furthermore, the number of words that avoid secondary structures was also shown to be exponentially large 
 by Milenkovic and Nashyap \cite{milenkovic2006design}.

\vspace{-0.1in}
\subsection{Prefix-Synchronized DNA Codes} \label{sec:prefix}

Thus far, we described how to construct address sequences that may serve as unique identifiers of the blocks they are associated with. 
We also pointed out that once such address sequences are identified, user information has to be encoded 
so as to \emph{avoid} the appearance of any of the addresses, sufficiently long substrings of the addresses, or substrings similar to the addresses in the resulting codewords. 

Specifically, for a fixed set $\A$ of address sequences of length $n$, 
we define the set $\C_\A(\ell)$ to be the set of sequences of length $\ell$ 
such that each sequence in $\C_\A(\ell)$ does not contain any string belonging to $\A$.
Therefore, by definition, when $\ell<n$, the set $\C_\A(\ell)$ is simply the set of strings of length $\ell$.
Our objective is then to design an efficient encoding algorithm (one-to-one mapping) 
to encode a set $\I$ of messages into $\C_\A(\ell)$.
For the sake of simplicity, we let $\I=\{0,1,2,\ldots, |\I|-1\}$ and
as is usual with constrained coding, we hope to maximize $|\I|$.

Clearly, $|\I|\le |\C_\A(\ell)|$ and hence, it is of interest to determine the size of $\C_\A(\ell)$.
In the case, when $\A$ is a set of mutually uncorrelated strings, Yazdi \etal  \cite{yazdi2015rewritable} 
proved the following theorem. 
\begin{thm} \label{thm:generating}
Suppose that $\A$ is a set of $M$ mutually uncorrelated sequences of length $n$ over the alphabet $\left\{ \mathtt{A,T,C,G}\right\}$. 
Define $F(z)=\sum_{\ell=0}^\infty |\C_\A(\ell)|z^n$. Then
\begin{equation}
F(z)=\frac{1}{1-4z+Mz^n}. \label{eq:sizeC} 
\end{equation}
\end{thm}

We make certain observations on \eqref{eq:sizeC}.
When $M$ is fixed, it is easy to show that $F(z)=1/(1-4z+Mz^n)$ has only one pole with radius less than one for sufficiently large $n$.
Furthermore, if $R^{-1}$ is the pole of $F$, 
we can show that $1/4 < R^{-1} <1/(4-\epsilon(n))$ with $\epsilon(n)=o(1)$. Here, the asymptotic is computed with respect to $n$.
In other words, for the case where $M$ is fixed, the size of $\C_\A(\ell)$ is at least $(4-\epsilon(n))^\ell(1-o(1))$ (here, asymptotic is computed with respect to $\ell$).

In the case where $\A$ contains a single address $\va$, 
Morita \etal{} proposed efficient encoding schemes into $\C_{\{\va\}}(\ell)$ 
in the context of prefix-synchronized codes \cite{morita1996construction}. 
Based on the scheme of Morita \etal{}, Yazdi \etal{} developed another encoding method 
that encodes messages into $\C_\A(\ell)$ where $\A$ contains more than one address.
In this scheme, Yazdi \etal assume that $\A$ is mutually uncorrelated and all sequences in $\A$ end with the same base,
which we assume without loss of generality to be $\dG$.
We then pick an address $\va\triangleq (a_1,a_2,\ldots,a_n)\in\A$ and 
define the following entities for $1\le i\le n-1$,
\begin{align*}
\bar{A}_i & =\left\{ \mathtt{A,C,T}\right\} \setminus\left\{ a_{i}\right\}, \\
\va^{(i)} & =(a_{1},a_2,\ldots, a_{i}).
\end{align*}

In addition, assume that the elements of $\bar{A}_{i}$ are arranged 
in increasing order, say using the lexicographical ordering $\mathtt{A\prec C\prec T}$. 
We subsequently use $\bar{a}_{i,j}$ to denote the $j$-th smallest element in $A_{i}$,
for $1\leq j\leq\left|\bar{A}_{i}\right|$. 
For example, if $\bar{A}_{i}=\left\{ \mathtt{C,T}\right\},$ 
then $\bar{a}_{i,1}=\mathtt{C}$ and $\bar{a}_{i,2}=\mathtt{T}.$ 

Next, we define a sequence of integers $G_{n,1},G_{n,2},\ldots$ that
satisfies the following recursive formula
\vspace{-0.05in}
\[
G_{n,\ell}=\begin{cases}
3^{\ell}, & 1\leq \ell<n,\\
\sum_{i=1}^{n-1}\left|\bar{A}_{i}\right|G_{n,\ell-i}, & \ell \geq n.
\end{cases}
\vspace{-0.1in}
\]

For an integer $\ell \geq0$ and $y<3^{\ell}$, let $\theta_{\ell}\left(y\right)=\left\{ \mathtt{A,T,C}\right\} ^{\ell}$
be a length-$\ell$ ternary representation of $y$. Conversely, for each
$W\in\left\{ \mathtt{A,T,C}\right\} ^{\ell}$, let $\theta^{-1}\left(W\right)$
be the integer $y$ such that $\theta_{\ell}\left(y\right)=W.$ 
We proceed to describe how to map every integer $\{0,1,\ldots,G_{n,\ell}-1\}$ into 
a sequence of length $\ell$ in $\C_\A(\ell)$ and vice versa. 
We denote these functions as $\enc_{\va,\ell}$ and $\dec$, respectively.

The steps of the encoding and decoding procedures are listed in Algorithm~\ref{table:alg} and the
correctness of was demonstrated by Yazdi \etal.

\begin{thm}
Let $\A$ be a set of mutually uncorrelated sequences that ends with the same base.
Then $\enc_{\va,\ell}$ is an one-to-one map from $\{0,1,\ldots,G_{n,\ell}-1\}$ to $\C_\A(\ell)$
and for all $x\in\{0,1,\ldots,G_{n,\ell}-1\}$, $\dec_\va(\enc_{\va,\ell}(x))=x$.
\end{thm}

In their experiment, Yazdi \etal{} found a set $\A$ of $M=32$ address sequences of length $n=20$ and 
used this method to encode information into $\C_\A(\ell=80)$. 
In this instance, the value of $G_{20,80}=1.56\times 10^{38} \ge 126$ bits, 
while the size of $\C_\A(80)$ is $1.462 \times 10^{48}\ge 159$ bits. 

The previously described 
$\enc_{\va,\ell}(x)$ algorithm imposes
no limitations on the length of a prefix used for encoding. This feature
may lead to unwanted cross hybridization between address primers used
for selection and the prefixes of addresses encoding the information. One approach
to mitigate this problem is to \textquotedblleft{}perturb\textquotedblright{} long prefixes in the encoded information 
in a controlled manner. For small-scale random access/rewriting experiments, the recommended approach is to
first select all prefixes of length greater than some predefined threshold.
Afterwards, the first and last quarter of the bases of these long prefixes
are used unchanged while the central portion of the prefix string
is cyclically shifted by half of its length.

For example, for the address primer $\va=\mathtt{ACTAACTGTGCGACTGATGC}$,
if the prefix $\va^{(16)}=\mathtt{ACTAACTGTGCGACTG}$ 
appears as a subword, say $\vp$, in $X=\enc_{\va,\ell}(x)$ 
then $X$ is modified to $X^{\prime}$ by mapping $\vp$ to $\vp^{\prime}=\mathtt{ACTAATGCCTGGACTG}$. 
This process of shifting is illustrated below:

\[
\begin{array}{cccc}
X & =\ldots & \overset{\vp}{\overbrace{\mathtt{ACTGT}\underbrace{\mathtt{GCGACT}}\mathtt{GATGC}}} & \ldots\\
\Downarrow &  & \overset{\textrm{cyclically shift by 3}}{\Downarrow}\\
X^{\prime} & =\ldots & \underset{\vp^{\prime}}{\underbrace{\mathtt{ACTGT}\overbrace{\mathtt{ACTGCG}}\mathtt{GATGC}}} & \ldots
\end{array}
\]

For an arbitrary choice of the addresses, this scheme may not allow for 
unique decoding $\enc_{\va,\ell}(x)$. 
However, there exist simple conditions 
that can be checked to eliminate primers that do not allow this transform to be ``unique''. 
Given the address primers created for our random access/rewriting experiments, we were able to uniquely map 
each modified prefix to its original prefix and therefore uniquely decode the readouts.

As a final remark, we would like to point out that prefix-synchronized coding also supports error detection and 
limited error-correction. Error-correction is achieved by checking if each substring of the sequence represents a 
prefix or ``shifted'' prefix of the given address sequence and making proper changes when needed. 

\begin{table*}[ht]
\begin{tabular*}{\hsize}
{@{\extracolsep{\fill}}rrrrrrr}
\hline
\multicolumn{1}{l}{\textbf{Algorithm 1} Encoding and decoding}\cr
\hline
\multicolumn{1}{l}{$X=\enc_{\va,\ell}(x)$}&
\multicolumn{4}{l}{$x=\dec_\va\left(X\right)$}\cr
\multicolumn{1}{l}{begin}&
\multicolumn{4}{l}{begin}\cr
\multicolumn{1}{l}{1$\quad$ if $\left(\ell \geq n\right)$}&
\multicolumn{4}{l}{1$\quad$$\ell=\textrm{length }\left(X\right);$}\cr
\multicolumn{1}{l}{2$\quad\quad$$t\gets1;$}&
\multicolumn{4}{l}{2$\quad$$X=X_{1}X_{2}\ldots X_{\ell};$}\cr
\multicolumn{1}{l}{3$\quad\quad$$y\gets x;$}&
\multicolumn{4}{l}{3$\quad$if $\left(\ell<n\right)$}\cr
\multicolumn{1}{l}{4$\quad\quad$ while $\left(y\geq\left|\bar{A}_{t}\right|G_{n,\ell-t}\right)$}&
\multicolumn{4}{l}{4$\quad\quad$ return $\theta^{-1}\left(X\right);$}\cr
\multicolumn{1}{l}{5$\quad\quad\quad$$y\gets y-\left|\bar{A}_{t}\right|G_{n,\ell-t};$}&
\multicolumn{4}{l}{5$\quad$else}\cr
\multicolumn{1}{l}{6$\quad\quad\quad$$t\gets t+1;$}&
\multicolumn{4}{l}{6$\quad\quad$ find$(s,t)\textrm{ such that }\va^{(t-1)}\bar{a}_{t,s}=X_{1}\ldots X_{t};$}\cr
\multicolumn{1}{l}{7$\quad\quad$ end;}&
\multicolumn{4}{l}{7$\quad\quad$ return $\left(\sum_{i=1}^{t-1}|\bar{A}_{i}|G_{n,\ell-i}\right)+(s-1)G_{n,\ell-t}+\dec_{\va}(X_{t+1}\ldots X_{\ell});$}\cr
\multicolumn{1}{l}{8$\quad\quad$$a\gets \lfloor y/{G_{n,\ell-t}\rfloor}$;}&
\multicolumn{4}{l}{8$\quad$end;}\cr
\multicolumn{1}{l}{9$\quad\quad$$b\gets y \bmod{G_{n,\ell-t}}$;}&
\multicolumn{4}{l}{end;}\cr
\multicolumn{1}{l}{10$\quad\quad$return $\va^{(t-1)}\bar{a}_{t,a+1}\enc_{\va,\ell-t}(b);$}\cr
\multicolumn{1}{l}{11$\quad$ else}\cr
\multicolumn{1}{l}{12$\quad\quad$ return $\mathcal{\theta}_{\ell}\left(y\right);$}\cr
\multicolumn{1}{l}{13$\quad$ end;}\cr
\multicolumn{1}{l}{end;}\cr
\hline
\end{tabular*}
\label{table:alg}
\end{table*}

\subsection{Error-Control Coding for DNA Storage}

\begin{figure}
\begin{centering}
\includegraphics[scale=0.18]{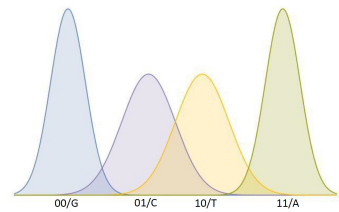} \label{fig:nanopores}
\vspace{-0.29in}
\par\end{centering}
\protect\caption{Impulse response of prototypical solid state nanopore sequencers.}
\vspace{-0.25in}
\end{figure}

Based on the discussion of error mechanisms in DNA synthesis and sequencing, it is apparent that most errors follow into the following categories:
\begin{itemize}
\item \emph{Substitution errors introduced during synthesis.} These errors may be addressed using many classical coding schemes, such as Reed-Solomon and Low-Density Parity-Check coding methods~\cite{gallager1962low}. One non-trivial problem associated with substitution errors introduced during the synthesis phase arises after high-throughput sequencing. In this case, errors in the synthesized sequences propagate through a number of reads produced during sequencing, and hence correspond to a previously unknown class of burst errors. The authors addressed this issue in a companion paper~\cite{kiah2015codes, kiah2014codes}, where they introduced the notion of \emph{DNA profile codes}, which have the property that they can correct combinations of sequencing and synthesis errors in reads, in addition to missing coverage (i.e., missing read errors).
\item \emph{Single deletion errors introduced during synthesis.} Isolated single deletion errors may be corrected by using Levenshtein-Tenengolts codes~\cite{varshamov1965code}, directly encoded into the DNA string. It also appears possible to extend the DNA profile coding paradigm to encompass deletion and insertion errors incurred during synthesis, although no results in this directions were reported.
\item \emph{Substitution and coverage errors introduced during sequencing.} These errors may be handled in a similar manner as substitution errors introduced during synthesis, provided that they are used with the correct sequencing platform (i.e., Illumina). For the third generation sequencing platforms - PacBio and Oxford Nanopore - only one specialized error-correction procedure was reported so far~\cite{gabrys2015asymmetric}, addressing problems arising due to overlapping impulse responses of two out of four bases (see Figure~\ref{fig:nanopores}).
\end{itemize}
It remains an open problem to design codes that efficiently combine all the constraints imposed by address design considerations and at the same provide robustness to both synthesis and sequencing errors.

\appendix  \label{sec:appendix}

\begin{itemize}
\item \textbf{Bases $A$, $T$, $G$ and $C$}: Nucleotides, the building units of DNA, include one out of four possible bases, {\tt A} (adenine), {\tt G} (guanine), {\tt C} (cytosine), and {\tt T} (thymine). With a slight abuse of meaning, we alternatively use the terms nucleotides and bases, and express DNA sequence lengths in nucleotides or basepairs.
\item \textbf{Capillary Electrophoresis}: Capillary electrophoresis is a technique that separates ions based on their electrophoretic mobility, observed when applying a controlled voltage. 
\item \textbf{Clone}: A section of DNA that has been inserted into a vector molecule, such as a plasmid, and then replicated to form many identical copies.
\item \textbf{Coverage (of a sequencing experiment)}: The average number of reads that contains a base at a particular position in the DNA string to be sequenced.
\item \textbf{De novo}: From scratch, without a template, anew.
\item \textbf{Deoxinucleotides}: Components of DNA, containing the phosphate,
sugar and organic base; when in the triphosphate form, they are the
precursors required by DNA polymerase for DNA synthesis (i.e., ATP,
CTP, GTP, TTP).
\item \textbf{DNA microarray}: A DNA microarray (also commonly known as
DNA chip or biochip) is a collection of microscopic DNA spots containing relatively short DNA
fragments termed probes, attached to a solid surface.
\item \textbf{DNA Hybridization}: DNA Hybridization is the process of combining
two complementary (in the Watson-Crick sense) single-stranded DNA or RNA molecules and allowing
them to form a single double-stranded molecule through base pairing.
\item \textbf{Dye-terminators}: Labeled versions of dideoxyribonucleotide triphosphates (ddNTPs), ``defective'' nucleotides used in Sanger sequencing.
\item \textbf{Enzyme}: Enzymes are biological molecules (proteins) that accelerate, or catalyze, chemical reactions.
\item \textbf{Heteroduplex}: A heteroduplex is a double-stranded (duplex)
molecule of nucleic acid originated through the genetic recombination
of single complementary strands derived from different sources, such
as from different homologous chromosomes or even from different organisms.
\item \textbf{Homologs}: Two chromosomes or fragments from chromosomes from a particular pair, containing the same genetic loci in the same order.
\item \textbf{Homopolymers}: Sequences of identical bases in DNA strings. 
\item \textbf{In vivo recombination}: Recombination is the process of combining genetic (DNA) material from multiple sources to create new sequences. In vivo recombination refers to recombination performed inside a living cell (in vivo).
\item \textbf{Ligase}: An enzyme that catalyzes the process of joining two molecules through the formation of new chemical bonds. 
\item \textbf{Luciferase}: An oxidative enzyme used to provide luminescence in natural or controlled
biological environments. 
\item \textbf{Oligonucleotide (short strand of nucleotides)}: A relatively short sequence of nucleotides, usually synthesized to match a region where a mutation is known to
occur.
\item \textbf{Polymerase chain reaction (PCR)}: Polymerase chain reaction (PCR)
is a laboratory technique used to amplify DNA sequences. The method
involves using short DNA sequences called primers to select the portion
of the genome to be amplified. The temperature of the sample is repeatedly
raised and lowered to help a DNA replication enzyme copy the target
DNA sequence. The technique can produce a billion copies of the target
sequence in just a few hours.
\item \textbf{Primer}: A primer is a strand of short nucleic acid sequences that serves as a starting point for DNA synthesis.
\item \textbf{Protein}: Proteins are large biological molecules, or macromolecules,
consisting of one or more long chains of amino acid residues.
\item \textbf{Read}: DNA fragment created during the sequencing process.
\item \textbf{Sequence assembly}: Sequence assembly refers
to aligning and merging fragments of a much longer DNA sequence in
order to reconstruct the original sequence.
\item \textbf{Symmetric dimer}: A chemical structure formed from two symmetric units.
\end{itemize}

\begin{figure}
\begin{centering}
\includegraphics[scale=0.55]{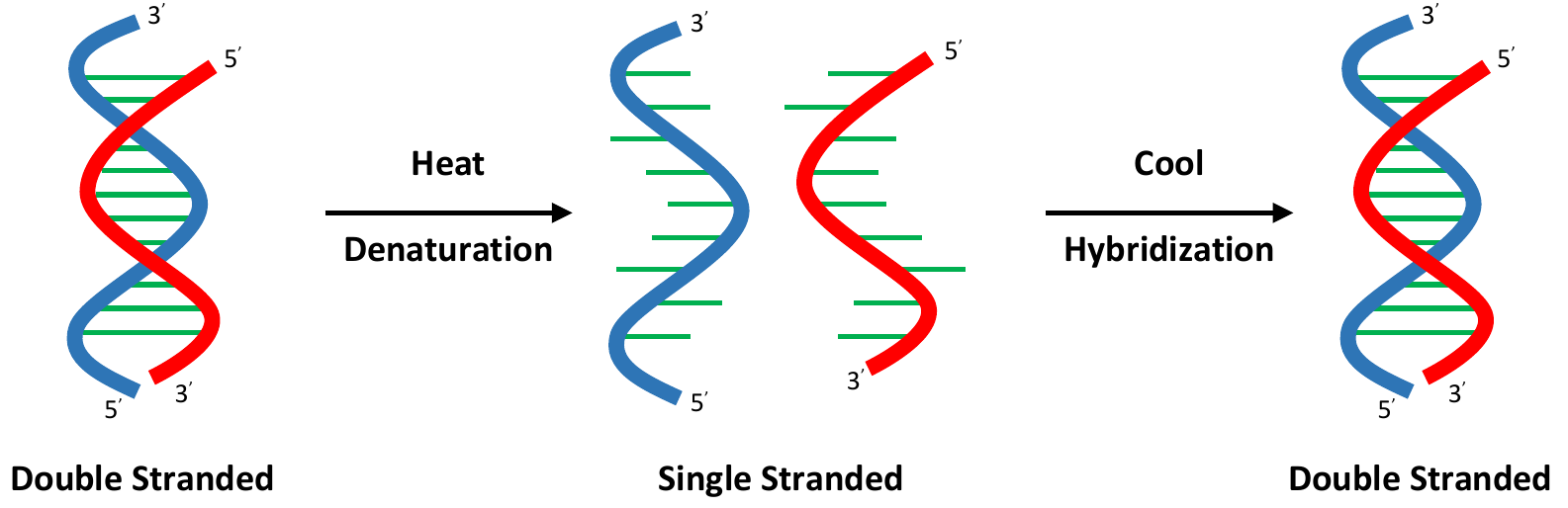} \label{fig:strategy}

\par\end{centering}
\protect\caption{Principles of DNA denaturation and hybridization.}
\vspace{-0.13in}
\end{figure}



\end{document}